\documentclass[ALICE,manyauthors]{cernphprep}

\usepackage[comma,square,numbers,sort&compress]{natbib}
\usepackage{hyperref}
\usepackage{booktabs}
\usepackage{lineno}
\usepackage{color}
\usepackage{multirow}
\usepackage[T1]{fontenc}
\usepackage{orcidlink}
\usepackage{textcomp}
\usepackage{placeins}

\newcommand{\nbar}{\ensuremath{{\rm \overline{n}}}\xspace}

\newcommand{\psigpbar}{\ensuremath{\overline{\Sigma}^{+}}\xspace}
\newcommand{\psigzbar}{\ensuremath{\overline{\Sigma}^{0}}\xspace}
\newcommand{\psigmbar}{\ensuremath{\overline{\Sigma}^{-}}\xspace}
\newcommand{\psigpmbar}{\ensuremath{\overline{\Sigma}^{\pm}}\xspace}

\newcommand{\pt}{\ensuremath{p_{\rm{T}}}\xspace}

\newcommand{\RpPb}{\ensuremath{R_{\rm{pPb}}}\xspace}

\newcommand{\ppb}{\ensuremath{\text{p--Pb}}\xspace}
\newcommand{\pp}{\ensuremath{\text{pp}}\xspace}
\newcommand{\gevc}         {Ge\kern-.1emV/$c$\xspace}
\newcommand{\mevc}         {Me\kern-.1emV/$c$\xspace}
\newcommand{\tev}          {Te\kern-.1emV\xspace}
\newcommand{\gev}          {Ge\kern-.1emV\xspace}
\newcommand{\mev}          {Me\kern-.1emV\xspace}
\newcommand{\gevmass}      {Ge\kern-.2emV/$c^2$\xspace}
\newcommand{\mevmass}      {Me\kern-.2emV/$c^2$\xspace}
\newcommand{\gevcsq}      {Ge\kern-.2emV/$c^2$\xspace}
\newcommand{\mevcsq}      {Me\kern-.2emV/$c^2$\xspace}

\newcommand{\snn}{\ensuremath{\sqrt{s_{\rm{NN}}}}\xspace}

\ifdefined\Ref
  \renewcommand{\Ref}[1]{Ref.\cite{#1}\xspace}
\else
  \newcommand{\Ref}[1]{Ref.\cite{#1}\xspace}
\fi

\begin{document}%

\begin{titlepage}
 \PHyear{2025}
 \PHnumber{151}      
 \PHdate{03 July}  
 %

  \title{\psigpmbar production in \pp and \ppb collisions at $\sqrt{s_{\rm NN}} = 5.02$ TeV with ALICE}
  \ShortTitle{\psigpmbar with PHOS}   
  \Collaboration{ALICE Collaboration\thanks{See Appendix~\ref{app:collab} for the list of collaboration members}}
  \ShortAuthor{ALICE Collaboration} 

  The transverse momentum spectra and integrated yields of anti-$\Sigma$ hyperons ($\psigpmbar$) have been measured in \pp and \ppb collisions at $\sqrt{s_{\mathrm{NN}}}=5.02$ TeV with the ALICE experiment.
Measurements are performed via the newly accessed decay channel $\psigpmbar \rightarrow \nbar\pi^{\pm}$. A new method of antineutron reconstruction with the PHOS electromagnetic spectrometer is developed and applied to this analysis.
The \pt spectra of $\psigpmbar$ are measured in the range $0.5<\pt<3$ \gevc and compared to predictions of the PYTHIA~8, DPMJET, PHOJET, EPOS LHC and EPOS4 models. The EPOS LHC and EPOS4 models provide the best descriptions of the measured spectra both in \pp and \ppb collisions, 
while models which do not account for multiparton interactions provide a considerably worse description at high \pt. 
The total yields of $\psigpmbar$ in both \pp and \ppb collisions are compared to predictions of the  Thermal-FIST model and dynamical models PYTHIA~8, DPMJET, PHOJET, EPOS LHC and EPOS4. All models reproduce the total yields in both colliding systems within uncertainties.
The nuclear modification factors $R_\mathrm{pPb}$ for both $\psigpbar$ and $\psigmbar$ are evaluated and compared to those of protons, $\Lambda$ and $\Xi$ hyperons, and predictions of EPOS LHC and EPOS4 models. No deviations of $R_\mathrm{pPb}$ for $\psigpmbar$ from the model predictions or measurements for other hadrons are found within uncertainties.

\end{titlepage}
\setcounter{page}{2}

\section{Introduction} \label{sec:intro}
Enhancement of strangeness production was one of the first proposed signatures of quark--gluon plasma  (QGP) formation in heavy-ion collisions~\cite{Rafelski}.
Many measurements have been performed since then, and
this idea has dramatically evolved.
Strangeness enhancement was first observed at SPS by experiments WA97~\cite{WA97:1999uwz} and NA57~\cite{NA57:2010tnk}, then it was seen also at RHIC~\cite{PHENIX:2004vcz,STAR:2008med} and at LHC energies~\cite{ALICE:2013xmt,CMS:2016zzh}.
Unexpectedly, an enhancement of strange hadron production was observed not only in nucleus-nucleus (AA) collisions but also in p--A (proton-nucleus) and even in high-multiplicity proton-proton (\pp) collisions~\cite{ALICE:2017jyt, ALICE:2013xmt, ALICE:2022wpn, Adam:2015vsf}, exhibiting a smooth transition from \pp to AA collisions as a function of the charged-particle density.

These observations stress the importance of the details of strangeness production, especially in \pp and p--A collisions. 
$\Sigma$ hyperons are of special interest as on the one hand they have minor feed-down of heavier resonance decays~\cite{ParticleDataGroup:2024cfk} and allow for direct tests of strangeness production, and on the other hand $\Sigma^{0}$ produces a considerable contribution to the $\Lambda$ yield.
$\Sigma$ antihyperons contain a single strange antiquark and form a triplet, with the electric charge defined by their light quark content
$\psigpbar$($\bar{d}\bar{d}\bar{s}$), $\psigzbar$($\bar{u}\bar{d}\bar{s}$) and $\psigmbar$($\bar{u}\bar{u}\bar{s}$). 
Both charged \psigpmbar have similar masses ($1189.37 \pm 0.07$ MeV for $\psigmbar$ and $1197.449 \pm 0.029$ MeV for $\psigpbar$) while their mean lifetimes are significantly different ($(8.018 \pm 0.026) \times 10^{-11}$ s for $\psigmbar$ and $(1.479 \pm 0.011) \times 10^{-10}$ s for $\psigpbar$)~\cite{ParticleDataGroup:2024cfk}.

$\Sigma$-baryon production was studied in detail in $e^+e^-$ collisions via Z hadronic decays by the experiments DELPHI~\cite{DELPHI:1995dso}, L3~\cite{L3:2000wwp}, and OPAL~\cite{OPAL:1996dbo} at LEP.
Theoretical models, such as the thermal statistical model~\cite{Andronic:2008ev}, describe these results within uncertainties.
However, there is no experimental data on $\Sigma$ hyperon production in hadron collisions, and  
it was not investigated whether those models can also describe $\Sigma$ production in these collisions.
Studying $\psigpmbar$-baryon production allows one to test different theoretical models and get insights into strange-baryon production mechanisms. Comparison of strange hadron spectra in pp and p--Pb collisions is important to understand the impact of multiparton interactions and their participation in the collective expansion of hot matter.

Hyperons provide an excellent opportunity to study the spin alignment ("global polarization") of final hadrons. The polarization of final hadrons may be produced via spin-orbit coupling by the specific velocity and vorticity fields  developed in the course of the evolution of hot matter created in AA collisions~\cite{Liang:2004ph}.
Global polarization of $\Lambda$ hyperons was observed in Pb--Pb collisions~\cite{ALICE:2021pzu,ALICE:2019onw}. These results can be described by hydrodynamic models. However, similar hydrodynamic models failed to reproduce global $\Lambda$ polarization observed in \ppb collisions~\cite{CMS:2025nqr}. The $\Sigma^0$ hyperons provide considerable feed-down for the $\Lambda$ and therefore, it is crucial to constrain this contribution, through the related $\Sigma^\pm$ particles, to make a quantitative comparison to the models. 

For the first time, $\psigpmbar$ are measured in the channels $\psigpbar \to \nbar \pi^{+}$ (BR = 99.848\%) and $\psigmbar \to \nbar \pi^{-}$ (BR = 48.31\%).
The \nbar identification technique and the measurement of its momentum with good accuracy pave the way for further measurements including \nbar production yields and correlations with other particles to test hadron formation mechanisms. In this paper, the \psigpmbar production was studied by employing a new analysis method in the field of high-energy physics, which relies on the possibility of reconstructing an antineutron (\nbar) in an electromagnetic calorimeter.

\section{Experimental apparatus, data sample and analysis} \label{sec:analysis}
The ALICE experiment was designed to explore the QGP formed in ultrarelativistic AA collisions, but its scope also covers physics in smaller systems such as pp and p--A  collisions.
The detector is optimized to provide excellent tracking at low \pt and particle identification over a wide range of momentum with various subdetectors.
To reconstruct \nbar, the Photon Spectrometer (PHOS)~\cite{Dellacasa:1999kd} is used while $\pi^\pm$ are reconstructed and identified with the central tracking system consisting of the Inner Tracking System~(ITS)~\cite{Aamodt:2010aa} and the Time Projection Chamber~(TPC)~\cite{Alme:2010ke}.

The PHOS is a precise electromagnetic calorimeter based on PbWO$_4$ crystals.
It is installed at a distance of 4.6~m from the nominal Interaction Point (IP), covering $70^\circ$ in azimuthal angle and $|\eta|<0.125$ in pseudorapidity.
The PbWO$_4$ crystals have the size of $2.2\times 2.2\times 18$ cm$^{3}$, where transverse dimensions were chosen close to the Moli\`ere radius. This ensures high granularity and provides the possibility to reconstruct showers in several cells to distinguish electromagnetic and hadronic showers using shower shape analysis.
The longitudinal size corresponds to 20 radiation length $X_{0}$ 
and $\sim 1$ interaction length $\lambda_\mathrm{int}$.

The ITS~\cite{Aamodt:2010aa} consists of six concentric silicon layers based on the silicon pixel (SPD), silicon drift (SDD), and silicon strip (SSD) technologies.
The TPC~\cite{Alme:2010ke} is a large cylindrical drift detector providing a maximum of 159 reconstructed space points per track and particle identification via the measurement of the specific energy loss d$E$/d$x$.
The central barrel detectors are installed in a magnetic field of $B=0.5$~T generated by a solenoid magnet.
The ITS covers a pseudorapidity range of $|\eta|<1.2$ while the TPC provides the reconstruction of tracks with good quality within $|\eta|<0.9$.
The ITS and TPC have full azimuthal coverage around the IP and are used for track reconstruction in the analysis.

The minimum bias~(MB) event trigger is organized similarly in \pp and \ppb collisions.
It is based on the coincidence of the signals of the V0A and V0C detectors consisting of two arrays of 32 scintillator tiles each covering the full azimuthal angle at $2.8<\eta<5.1$ and $-3.7<\eta<-1.7$, respectively~\cite{Abbas:2013taa}. Furthermore, the V0A and V0C detectors (together defined as V0M) are used for multiplicity measurements.

The \ppb data at $\snn=5.02$~TeV were recorded in 2016 with one beam configuration, where the circulation directions of the proton and Pb beams in LHC did not change. 
Equal magnetic rigidity for the proton and Pb beams in the LHC resulted in a rapidity shift of $\Delta y_{\rm NN}=-0.465$ in the direction of the proton beam between the nucleon--nucleon center-of-mass ($y_\mathrm{CMS}$) and the laboratory reference system.  
The \pp data at the same center of mass energy were collected in 2017.
For this analysis, inelastic (INEL) events were selected. 

\subsection{Antineutron identification}

The PHOS has a small hadronic thickness, and it is not able to reconstruct the full energy deposited by an \nbar.
In addition, the typical energy resolution of hadronic calorimeters $\sigma_\mathrm{E} \sim (30-70)\%/\sqrt{E}$~\cite{ParticleDataGroup:2024cfk} is
not sufficient to resolve a resonance peak in the high-multiplicity environment of a \ppb or even \pp collision.
Therefore, an alternative approach is used for the $\nbar$ identification and reconstruction: the $\nbar$ identification is performed using cluster properties, while the $\nbar$ momentum is estimated based on the timing information. 
A cluster in the PHOS consists of a set of cells (or PbWO$_4$ crystals) that share a common side or corner~\cite{Dellacasa:1999kd}. 
The clustering algorithm starts from the seed cell with the energy above the threshold $E_\mathrm{seed}=30$ MeV and adds all cells with the energy above minimal energy threshold $E_\mathrm{min}=10$ MeV and having a common side or corner with any cell already in the cluster. If several local maxima (cells with the energy larger than any adjacent cell has by $E_\mathrm{locMax}=30$ MeV) appear in the cluster, an unfolding algorithm is applied to split the cluster into several clusters~\cite{Dellacasa:1999kd}.

\begin{figure}[t!]
  \centering
  \includegraphics[width = 0.48\linewidth]{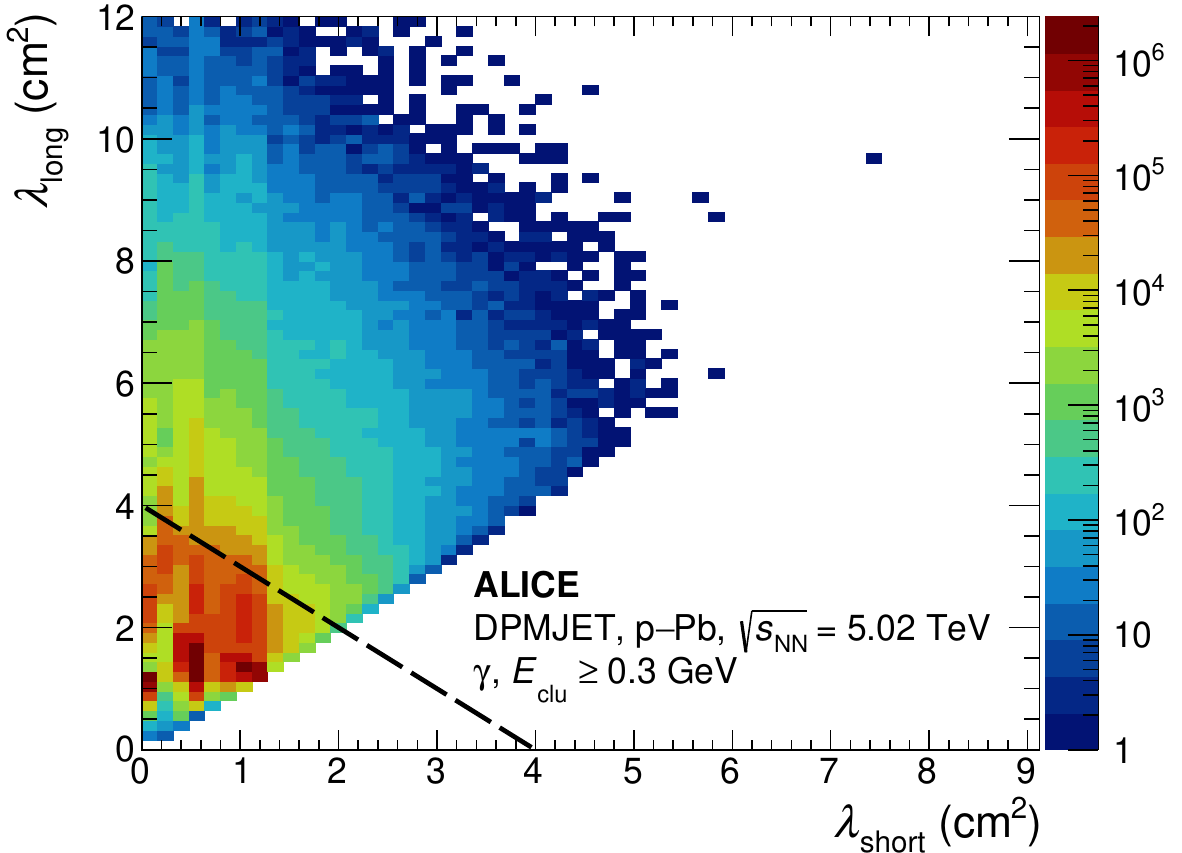}
  \hfill
  \includegraphics[width = 0.48\linewidth]{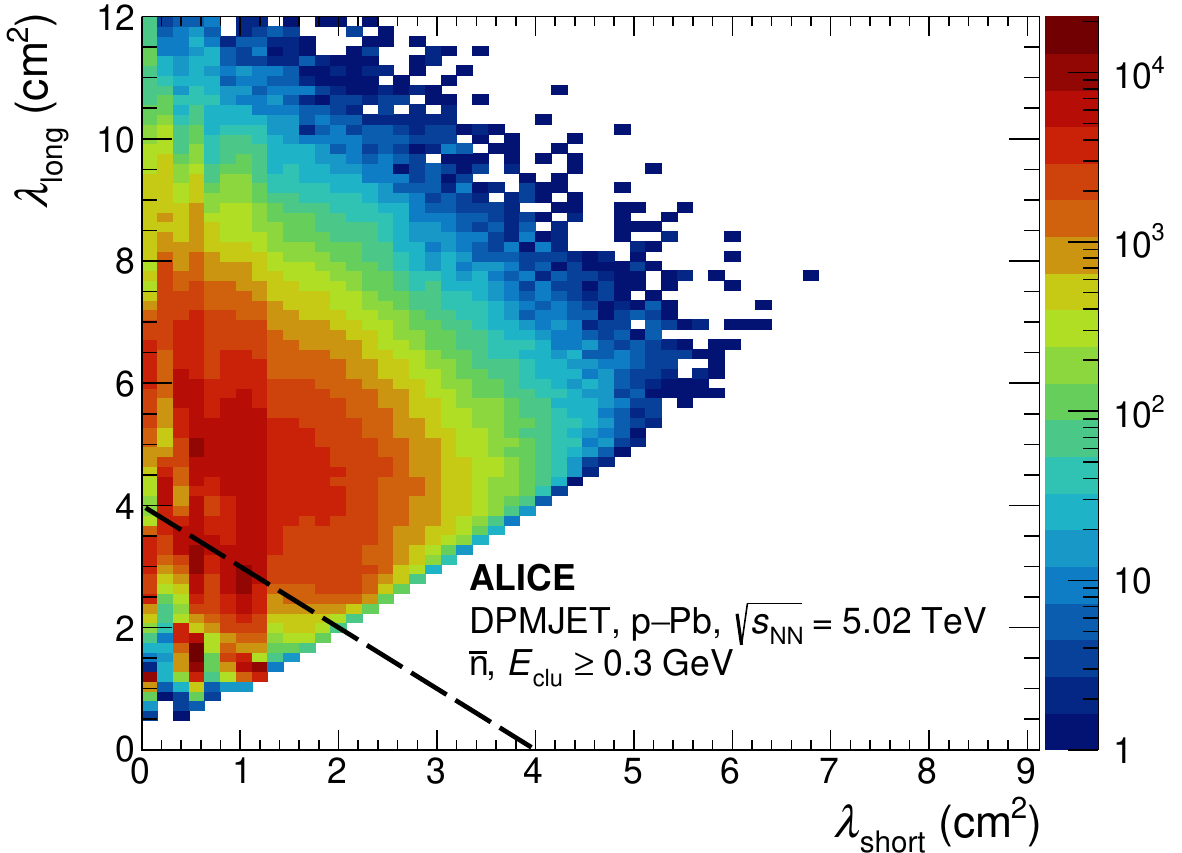}
  \caption{Shower shape parameters $\lambda_\mathrm{long}$ vs. $\lambda_\mathrm{short}$ for $\gamma$ (left) and ${\rm \overline{n}}$ (right) simulated in p--Pb collisions simulated with the DPMJET 3.0-5~\cite{dpmjet:2000he} event generator.}
  \label{fig:disp}
\end{figure}

The identification of \nbar clusters in the PHOS is based on three variables: the shower shape, the neutrality 
of a cluster, and the cluster energy.  
In addition, a loose time cut of 150 ns is applied for PHOS clusters to remove the pileup contribution.
The shower shape identification is based on the fact that showers produced by antibaryons are characterized by generally larger dispersion values compared to electromagnetic ones.
To quantify the shower shape, the two variables, $\lambda_\mathrm{short}$ and $\lambda_\mathrm{long}$, are used. They are by definition the eigenvalues of the two-dimensional dispersion matrix $M_\mathrm{ij}=\sum w_\mathrm{k}(x_\mathrm{i,k}-\bar{x_\mathrm{i}})(x_\mathrm{j,k}-\bar{x_\mathrm{j}})$ where weights $ w_\mathrm{k}= \max[0,4.5+\ln(E_\mathrm{k}/E_\mathrm{clu})]$~\cite{Dellacasa:1999kd}, $E_\mathrm{k}$ is the energy of the $k$-th cell of the cluster, $E_\mathrm{clu}$ is the cluster energy and $x_\mathrm{j,k}$ is the $j$-th coordinate in the calorimeter plane (along or perpendicular to the beam direction) of the $k$-th cell. 
Photons produce symmetric and compact clusters with $\lambda_\mathrm{short}\sim \lambda_\mathrm{long}\sim 1.5$~cm$^{2}$, while clusters produced by \nbar are generally larger and more asymmetric.
Fig.~\ref{fig:disp} shows $\lambda_\mathrm{short}$ and $\lambda_\mathrm{long}$ distributions obtained with Monte-Carlo simulations performed with the ALICE simulation and reconstruction framework AliRoot~\cite{ALICE:2005aa} and the DPMJET 3.0-5~\cite{dpmjet:2000he} event generator. The AliRoot framework includes a detailed description of the material of ALICE subdetectors, simulates responses of all detectors using GEANT 3 tracking, and applies a standard reconstruction procedure to the simulated data. 
Clusters are classified according to the particles depositing the largest part of the energy in the cluster.
A selection $\lambda_\mathrm{long}\geq a_\mathrm{disp}-\lambda_\mathrm{short}$ with $a_\mathrm{disp}=4$ cm$^{2}$, shown with dashed line on Fig.~\ref{fig:disp}, is applied in the analysis. Another possibility to distinguish an antineutron is to use the number of cells in a cluster. The mean number of cells in photon clusters with $E_\mathrm{clu}>0.5$~\gevc follows an approximately Poisson distribution with a mean of $\sim 2$ cells, while the number of cells in antineutron clusters has a wide distribution with a mean of $\sim 15$ cells.
This variable is correlated to the dispersion of a cluster, but an application of the selection $N_\mathrm{cell}> 7$ improves the purity of $\nbar$ sample.

Antiprotons may also induce hadronic showers with large dispersion values. The neutrality identification criterion is the absence of a track reconstructed in the central tracking system and extrapolated to the PHOS surface in the vicinity of a cluster (by default, $10\sigma$)~\cite{Blau:2020wda}. The $\sigma$ value here is the track $\pt$-dependent resolution of a track propagation to the PHOS. The resolution decreases with the track momentum and reaches approximately 1.4 cm along the beam and magnetic field direction and 2.4 cm in the perpendicular direction at a track $\pt=1$ \gevc. Antiprotons could have a large offset between the point of entry of a track to the PHOS and the center of gravity of a cluster, therefore, a relatively large veto selection is chosen.

\begin{figure}[t!]
  \centering
  \includegraphics[width = 0.49\linewidth]{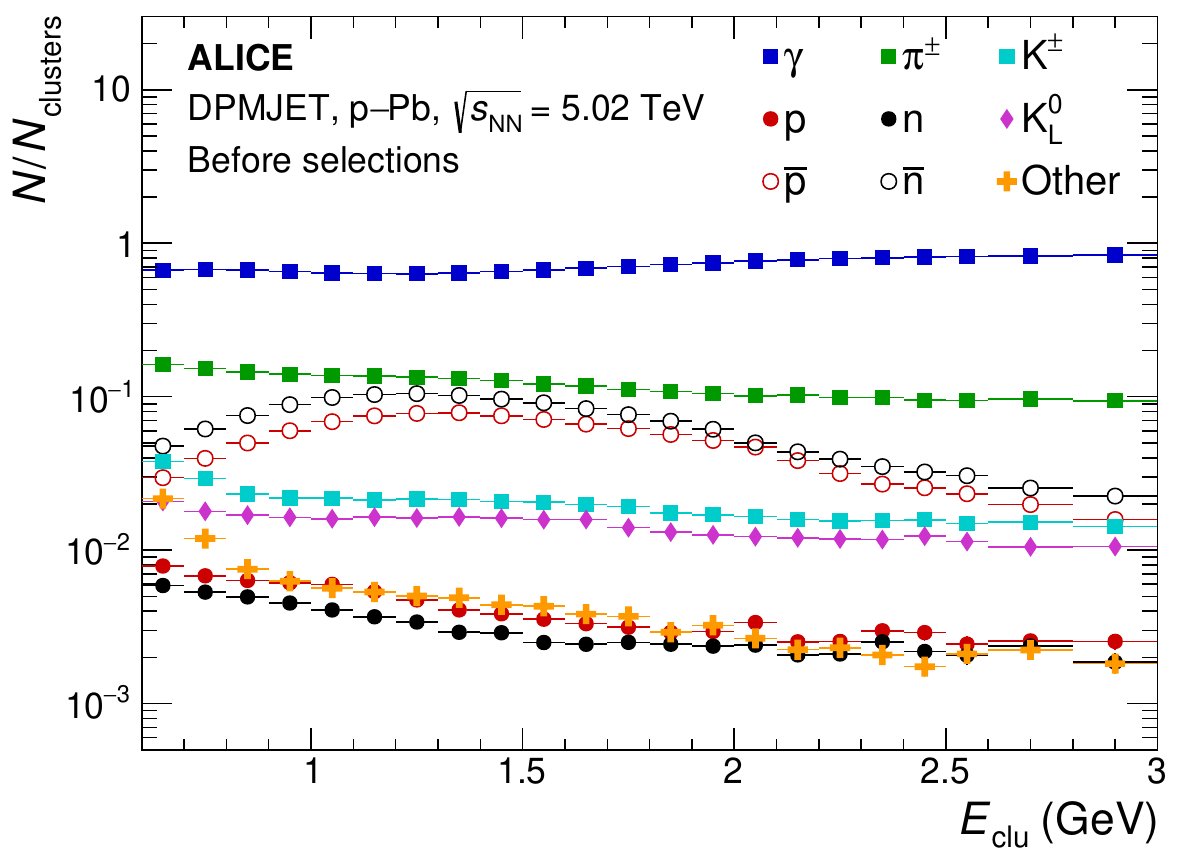}
  \hfill
  \includegraphics[width = 0.49\linewidth]{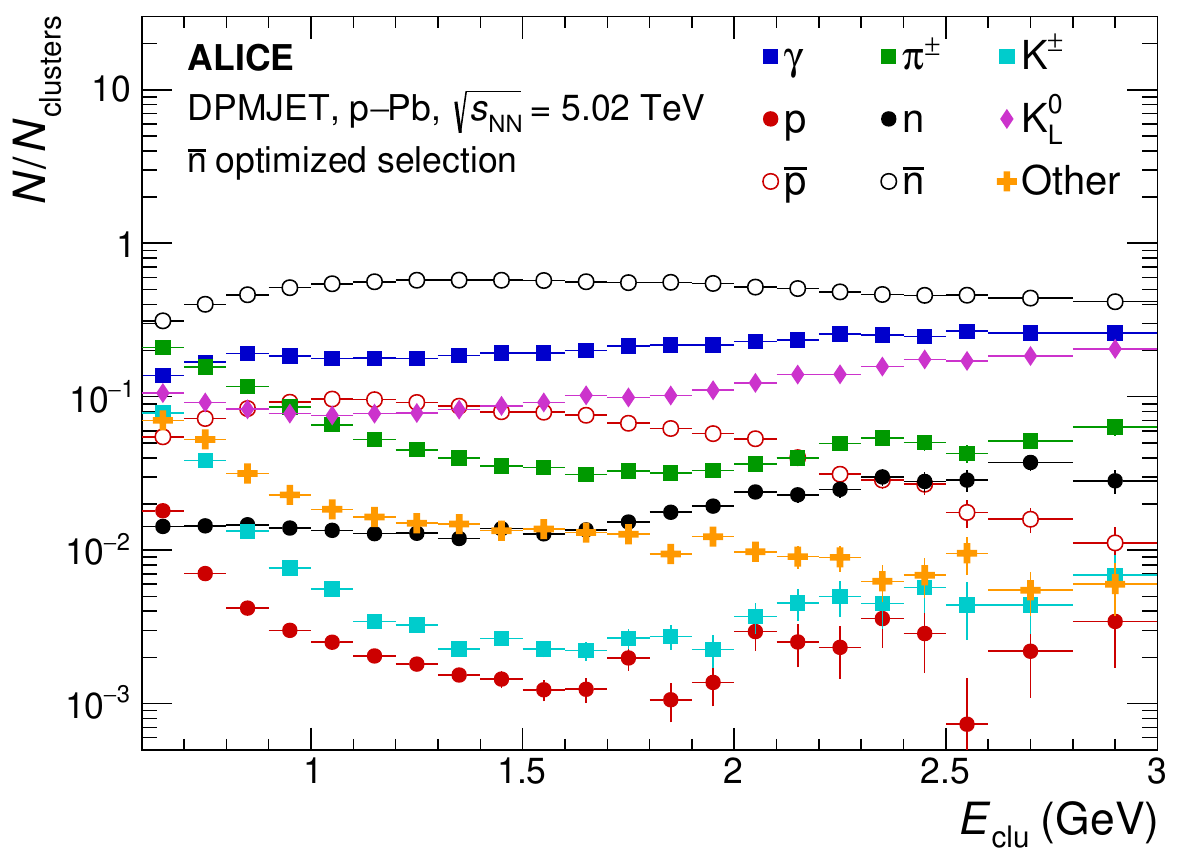}
  \caption{Fraction of different types of particles which produce clusters in PHOS before (left) and after the application of a default set of selections (right) as a function of cluster energy in p--Pb collisions simulated with the DPMJET 3.0-5~\cite{dpmjet:2000he} event generator.}
  \label{fig:pur}
\end{figure}

The spectrum of the PHOS clusters has a characteristic bump in the region $0.5<E_\mathrm{clu}<1.5$ GeV related to the (soft) antibaryon annihilation and the deposition of most of the produced energy in the calorimeter, see left column of the Fig.~\ref{fig:pur}. This figure shows the relative contributions of clusters produced by different particles in \ppb collisions as a function of the cluster energy. Before selections (left column), antibaryons contribute up to 10\% to the total cluster spectrum and produce a wide bump because of random hadron shower leakage. To increase the $\nbar$ contribution, a selection on a minimal cluster energy (by default, $E_\mathrm{clu}>0.5$~GeV) is applied.

Before the selections, the photons constitute the major fraction of all clusters, while by applying the selection criteria discussed above, one increases the fraction of \nbar to 50-60\% in the cluster energy range $0.5<E_\mathrm{clu}<2$~GeV, see Fig.~\ref{fig:pur}. The relative contributions of different particle species and improvement due to applied selections remain approximately the same in \pp and \ppb collisions.
Note that an energy boost due to an annihilation makes antineutron and antiproton clusters approximately 10 times more abundant than neutron and proton ones. Modest PHOS timing resolution of soft clusters makes an extension of this technique to neutrons very challenging.

The time of flight information $t$ extracted from PHOS clusters 
is used to calculate the \nbar momentum:
\begin{equation}
  p_{\text{rec}} = \frac{m_{n}c}{\sqrt{\left(c\cdot t/l\right)^2-1} },
  \label{eq:momentum}
\end{equation}
where $l$ is the distance between the primary vertex (PV) and a cluster in the PHOS and $m_{n}=0.9395$ GeV$/c^2$~\cite{ParticleDataGroup:2024cfk} is the neutron mass.
The PHOS timing resolution rapidly improves with the increase of a cluster energy $E_\mathrm{clu}$~\cite{ALICE:2019cox}. Therefore, a time measured in the cluster cell with the maximal energy is used.
To characterize the antineutron momentum resolution obtained with this method, a Full Width at Half Maximum (FWHM) of the reconstructed momentum distribution is estimated: FWHM=300 MeV/$c$ at transverse momentum $0.5$ GeV/$c$ and 500 MeV/$c$ at $p_\mathrm{T}=1.5$ GeV/$c$ which corresponds to resolutions of $125$ MeV/$c$ and 210 MeV/$c$, respectively, for Gaussian distributions. 

\subsection{Track selection, and decay topology}

To reconstruct charged pion tracks, the TPC and ITS were used. Only good quality tracks were selected within a pseudorapidity range of $|\eta|<0.8$ using the standard ALICE selection, requiring at least 50 associated hits in the TPC. To identify the $\pi^\pm$ tracks, a 3$\sigma$ selection on the difference between the measured and expected d$E$/d$x$ values in TPC for pions, normalized to the resolution,~\cite{ALICE:2014sbx} was imposed.

\begin{figure}[h!]
  \centering
  \includegraphics[width = 0.5\linewidth]{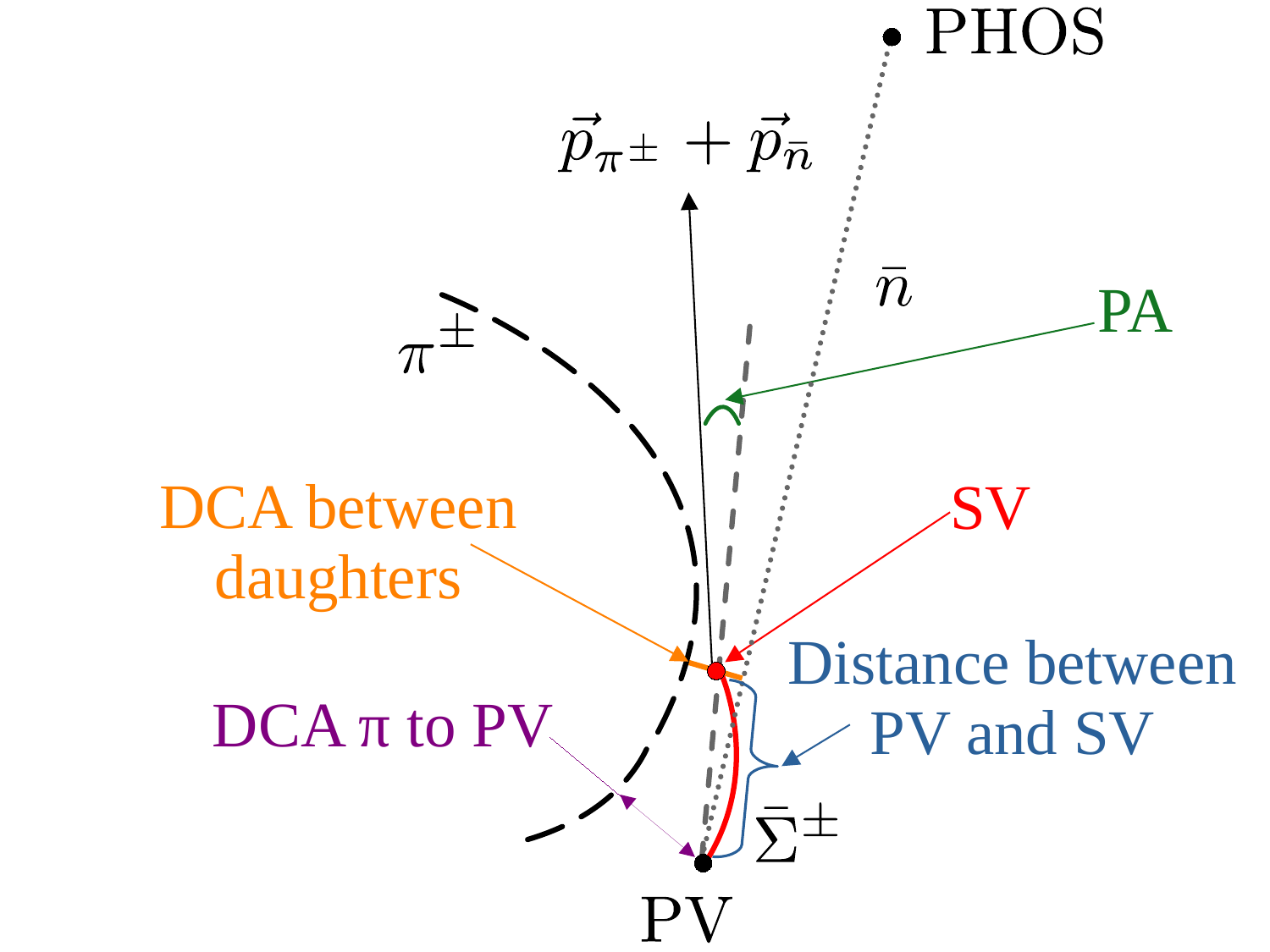}
  \caption{Decay topology of the $\overline{\Sigma}^{\pm} \to {\rm \overline{n}} \pi^\pm$ decay. In the reconstruction procedure, the direction of the ${\rm \overline{n}}$ momentum is approximated by the vector pointing from the the PV to the cluster in the PHOS. DCA of $\pi^\pm$ is not used in the analysis.}
\label{fig:topology}
\end{figure}

Due to the relatively large lifetimes of $\psigpmbar$, their decay vertices are shifted with respect to the primary vertex by a few centimetres so that one can construct and apply topological selections.
The secondary vertex (SV) is reconstructed as a point of the closest approach of the charged track to the straight line of the \nbar track connecting PV and a cluster in the PHOS. This approximation of the $\nbar$ track is viable because of the relation of $\psigpmbar$ lifetimes, the distance to the PHOS and due to $\psigpmbar$ decay kinematics.
The reconstruction of an SV allows one to use standard topological variables, see Fig.~\ref{fig:topology}: the Distance of Closest Approach (DCA) between the daughter particles in 3D space, the cosine of the Pointing Angle (CPA) between the direction of the total momentum and the direction of the vector pointing from the SV to the PV, and the decay radius. The default selections are summarized in Table~\ref{tab:top_default}.

\begin{table}[h!]
\centering
\caption{Default selections used in the analysis.}
\label{tab:top_default}
\begin{tabular}{|c|cc|}
\hline
Selection                   & \multicolumn{2}{c|}{Topology selections}                                                     \\ \hline
                            & \multicolumn{1}{c|}{$\overline{\Sigma}^+$}              & $\overline{\Sigma}^-$              \\ \hline
DCA$_{\text{daug}}$, cm     & \multicolumn{2}{c|}{$< 0.06 + \exp(-1.381\cdot p_{\text{T}} - 2.232)$}                       \\ \hline
CPA                         & \multicolumn{2}{c|}{$> 0.3$}                                                                 \\ \hline
Distance between PV and SV, cm & \multicolumn{1}{c|}{$> 0.193\cdot p_{\text{T}} + 0.25$} & $> 0.193\cdot p_{\text{T}} + 0.15$ \\ \hline
                            & \multicolumn{2}{c|}{Clusters}                                                                \\ \hline
Min $E_{\text{clu}}$, GeV   & \multicolumn{2}{c|}{$\geq 0.6$}                                                              \\ \hline
$N_{\text{cells}}$          & \multicolumn{2}{c|}{$\geq 7$}                                                                \\ \hline
Dispersion, cm$^2$          & \multicolumn{2}{c|}{$\lambda_{\text{long}} \geq -\lambda_{\text{short}}+4$}                  \\ \hline
CPV, $n_{\sigma}$           & \multicolumn{2}{c|}{$> 10$}                                                                  \\ \hline
                            & \multicolumn{2}{c|}{Tracks}                                                                  \\ \hline
TPC clusters                & \multicolumn{2}{c|}{$\geq 60$}                                                               \\ \hline
$|\eta|$                    & \multicolumn{2}{c|}{$\leq 0.8$}                                                              \\ \hline
TPC PID for $\pi$, $n_{\sigma}$ & \multicolumn{2}{c|}{$< 3$}                                                                   \\ \hline
\end{tabular}
\end{table}

\FloatBarrier

MC simulations show that the DCA and the CPA have a good selection power for the reconstruction of the $\psigpmbar$ decays. In contrast, the distance between the PV and the SV shows lower selection power. The usage of the $\pi^\pm$ DCA to the primary vertex is not increasing the signal to background ratio, thus it is not used in the analysis.

\subsection{Raw yield extraction and efficiency evaluation}

\begin{figure}[b]
  \centering
  \includegraphics[width = 0.48\linewidth]{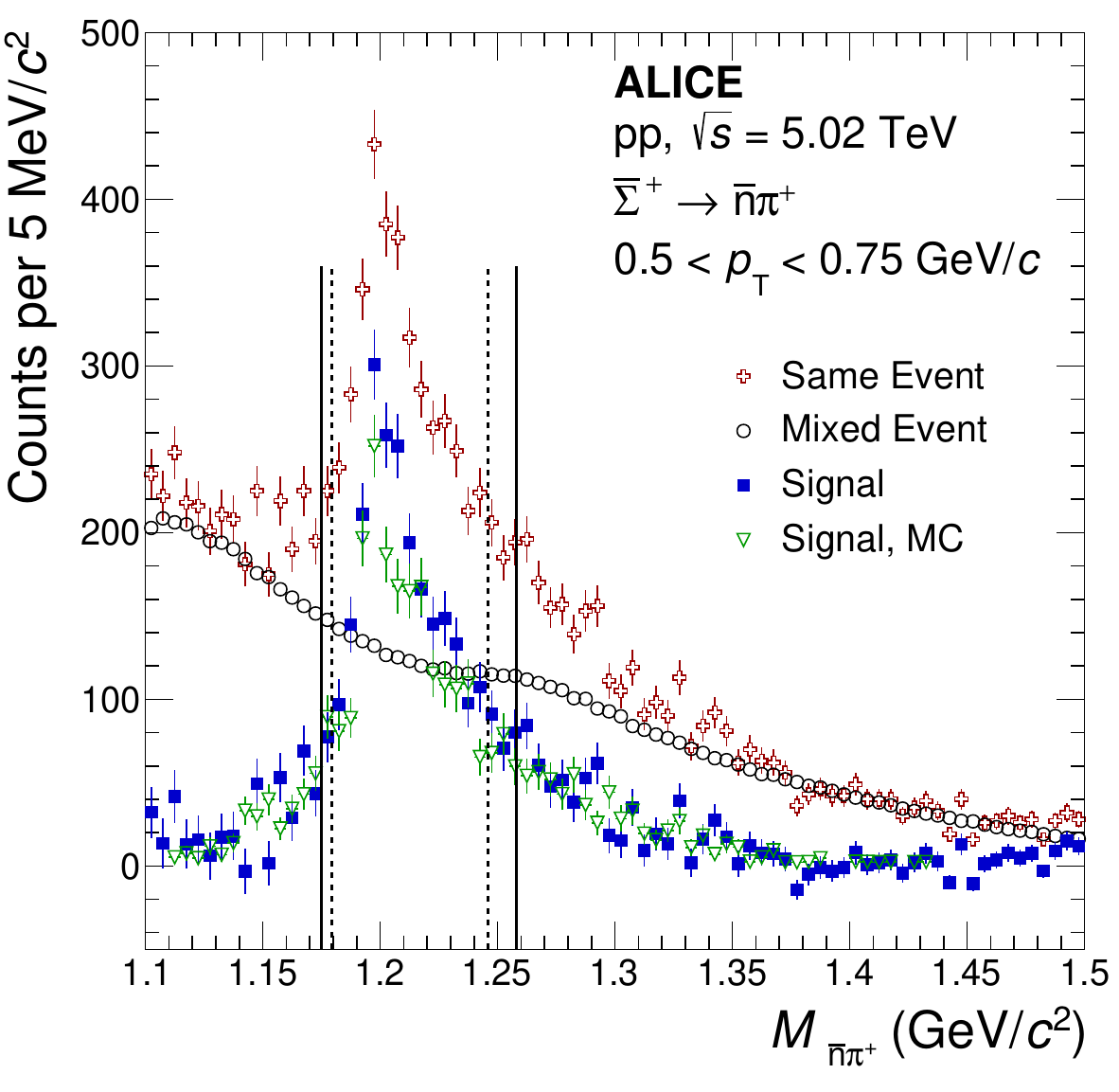}
  \hfill
  \includegraphics[width = 0.48\linewidth]{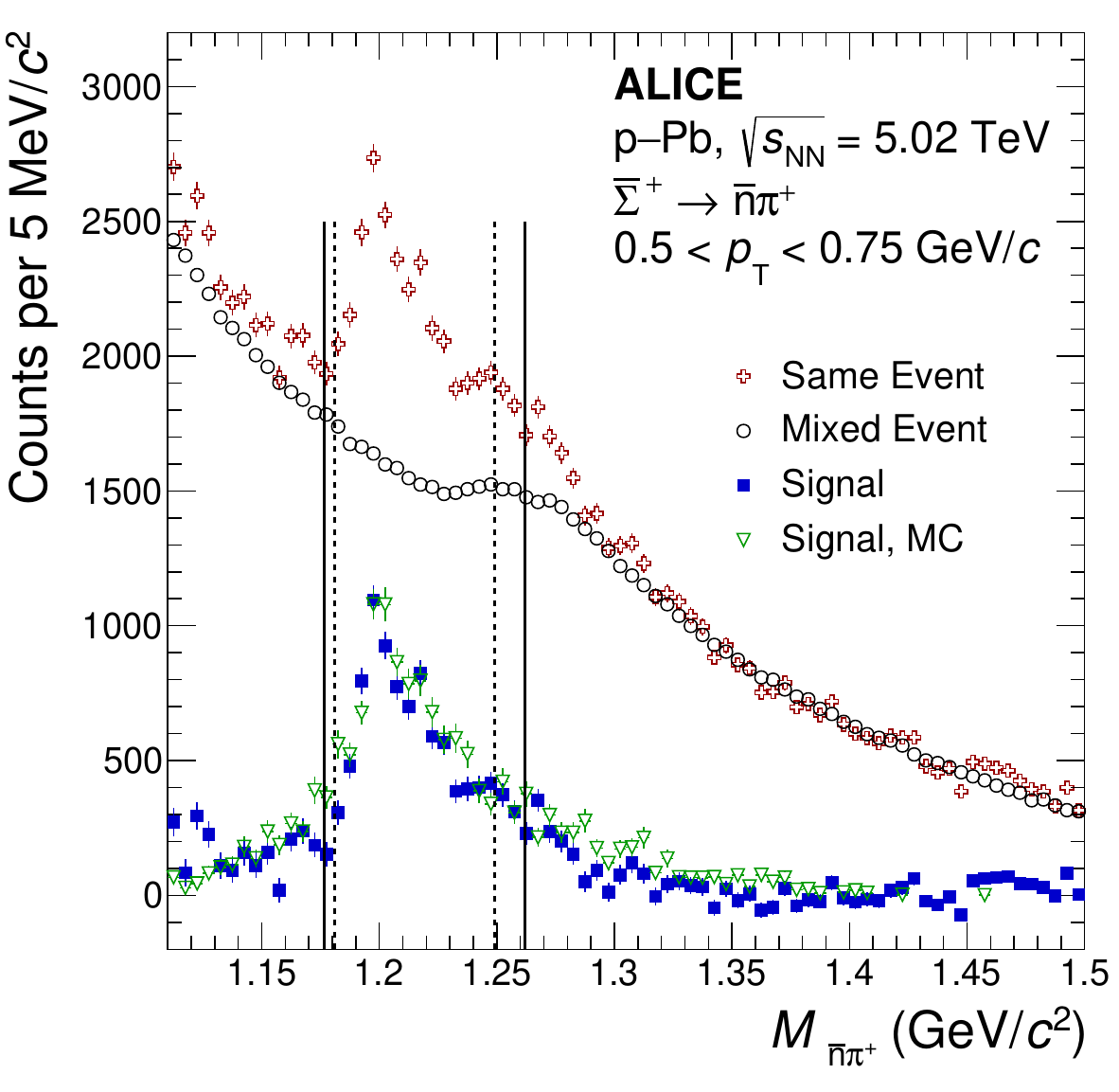}
  \hfill
  \includegraphics[width = 0.48\linewidth]{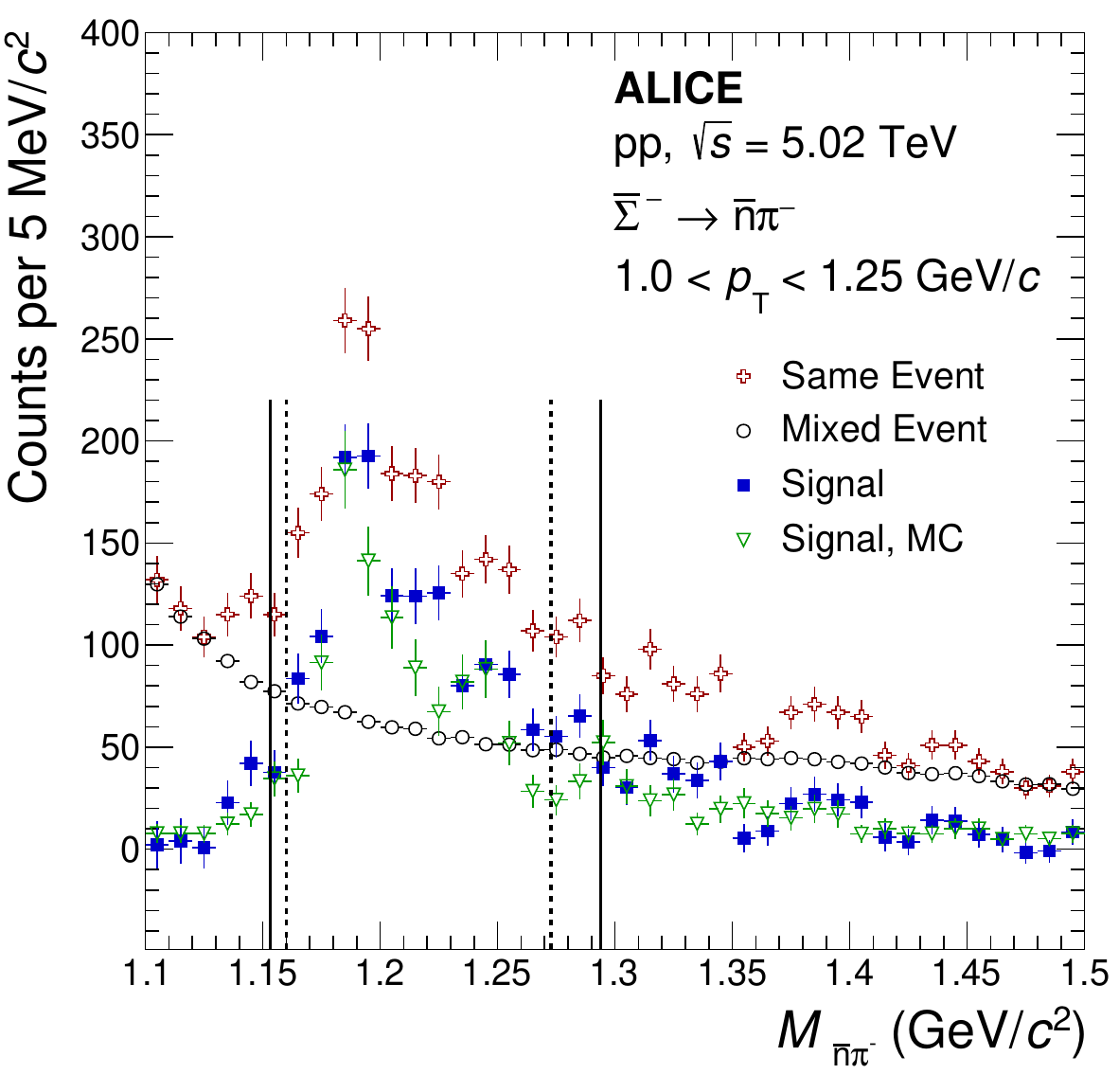}
  \hfill
  \includegraphics[width = 0.48\linewidth]{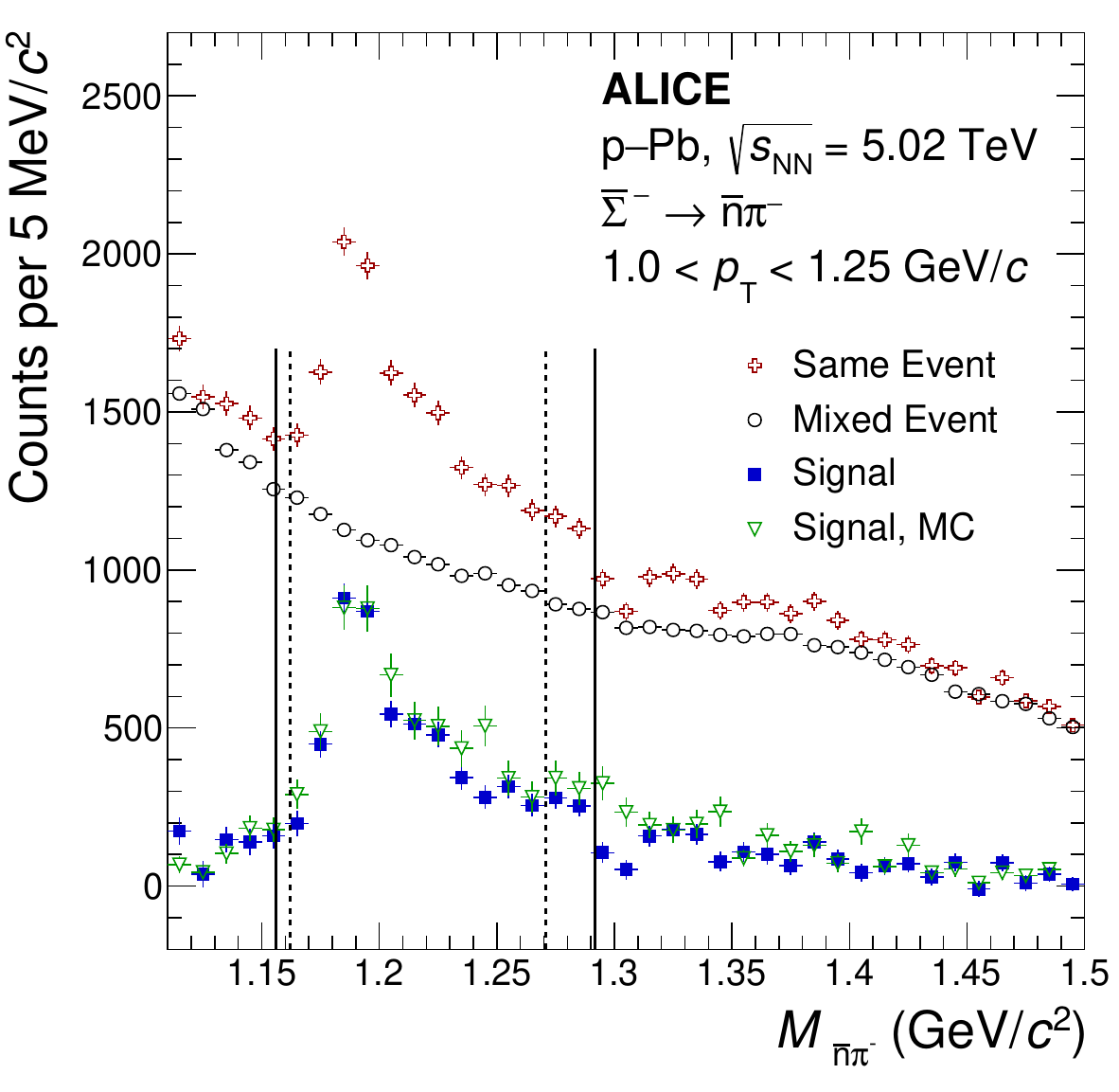}
  \caption{Same event, mixed event, signal from data, and MC true signal invariant mass distributions of ${\rm \overline{n}}\pi^+$ (top row) and ${\rm \overline{n}}\pi^-$ (bottom row) pairs in pp (left column) and p--Pb (right column) collisions for two selected $p_{\rm{T}}$ bins. Solid and dashed vertical lines represent integration ranges at 1/4 and 1/3 of the maximum, respectively. The PYTHIA~8~\cite{pythia} and DPMJET 3.0-5~\cite{dpmjet:2000he} event generators were used for Monte-Carlo simulations.}
  \label{fig:MpiN}
\end{figure}

The raw yields of $\psigpmbar$ are calculated using invariant mass distributions of $\nbar\pi^\pm$ pairs producing the same event (SE) distributions. Examples of such distributions are shown in Fig.~\ref{fig:MpiN}.
The particle identification and topological selections significantly increase the signal/background ratio of the $\psigpmbar$ peak and allow one to measure
the $\psigpmbar$ production down to $\pt=0.5$ \gevc.
To estimate and subtract the combinatorial background, an event mixing technique is used.
A sequence of 100 events with PV within 2 cm intervals is used for the mixing. In the case of \ppb collisions, the additional condition for mixed events to have the same 10\% multiplicity class is required.
Note the complicated shape of the combinatorial background, e.g. a bump at approximately $M_\mathrm{\nbar\pi^\pm}\sim1.27$ \gevcsq which is related to the limited PHOS acceptance, some inhomogeneity in the central tracking system acceptance and applied selection criteria. This shape is reproduced in the MC simulations.
To take into account possible residual correlations due to jets, collective flow, resonance decays, etc., the mixed event (ME) distribution is scaled with a linear function to reproduce the SE distribution beyond the peak. 
To estimate the scale function, a fit of SE/ME ratio with linear plus the peak function (see details below) in the range $1.1 < M_{\nbar\pi} < 1.5$ \gevcsq is performed. 
The signal distribution is calculated as a difference between the SE and scaled ME distributions.

\begin{figure}[b!]
  \includegraphics[width=.475\textwidth]{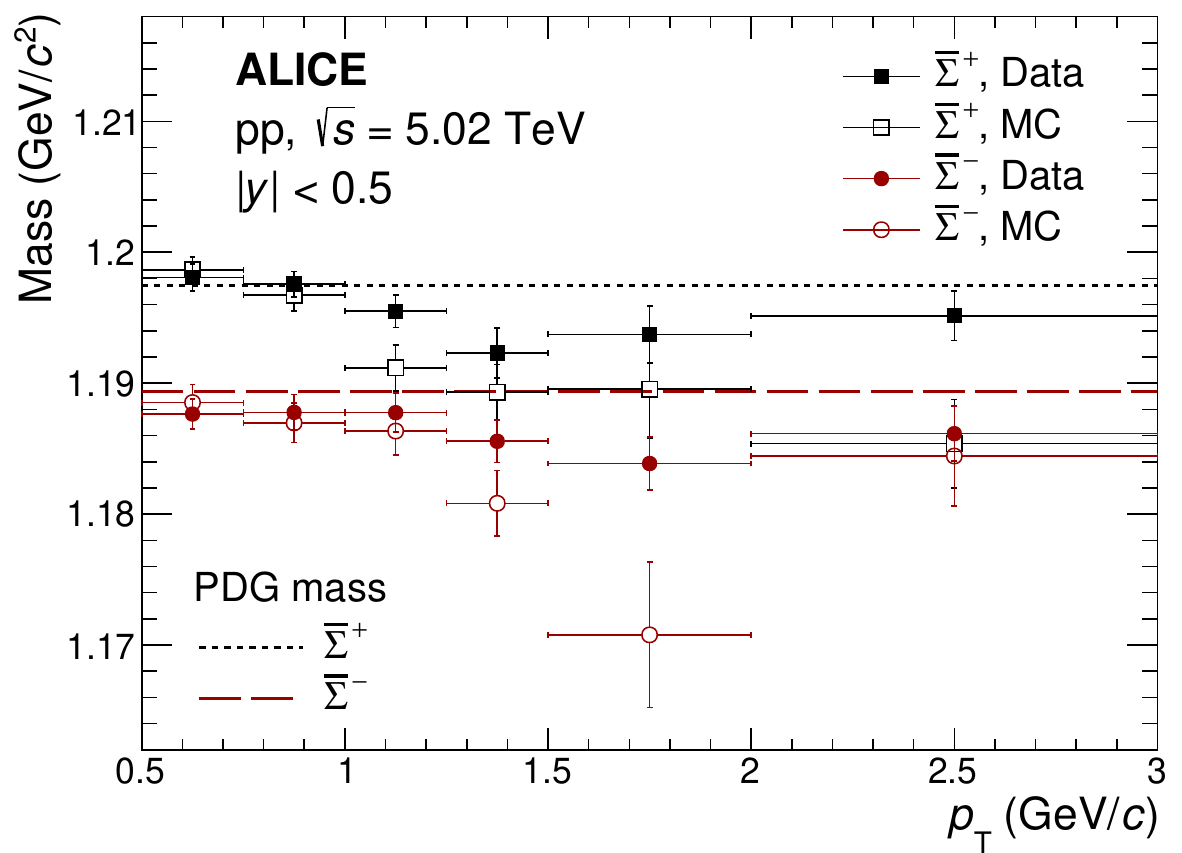}
  \hfill
  \includegraphics[width=.475\textwidth]{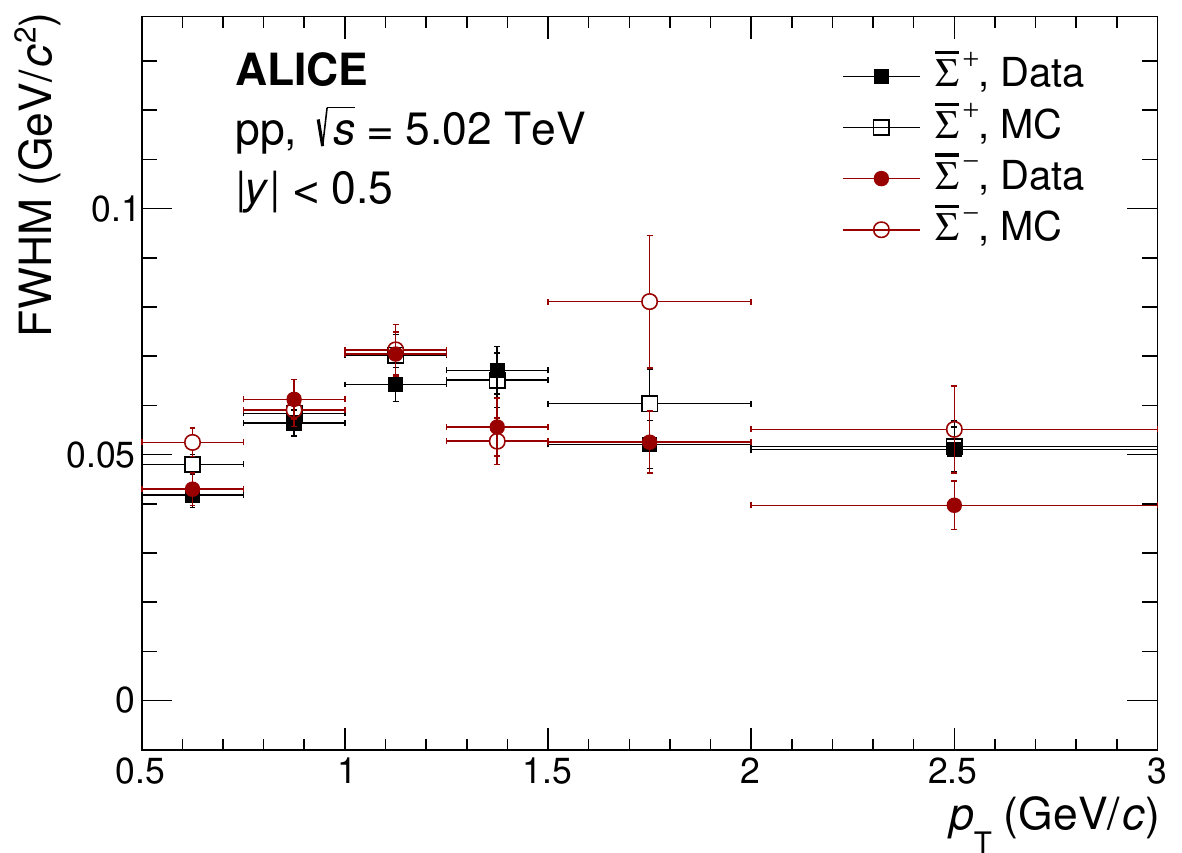}  \\
  \includegraphics[width=.475\textwidth]{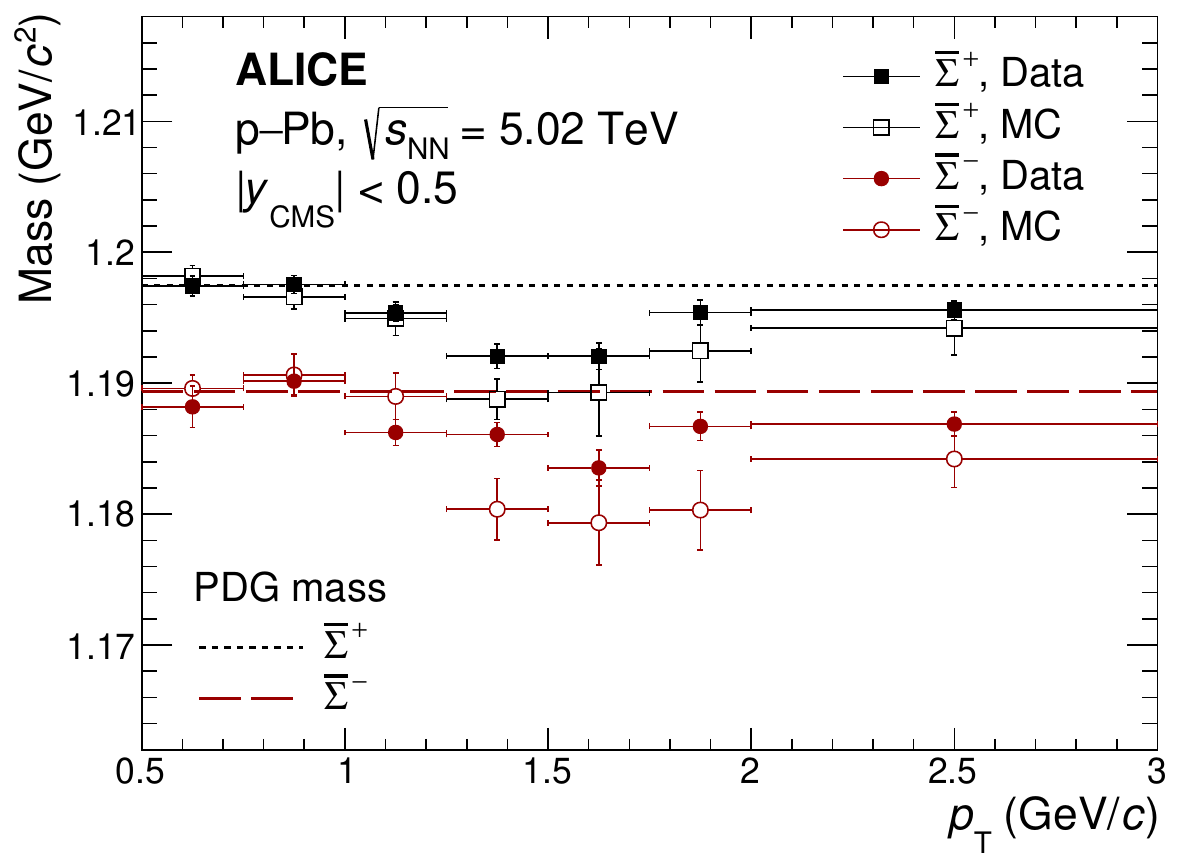}
  \hfill
  \includegraphics[width=.475\textwidth]{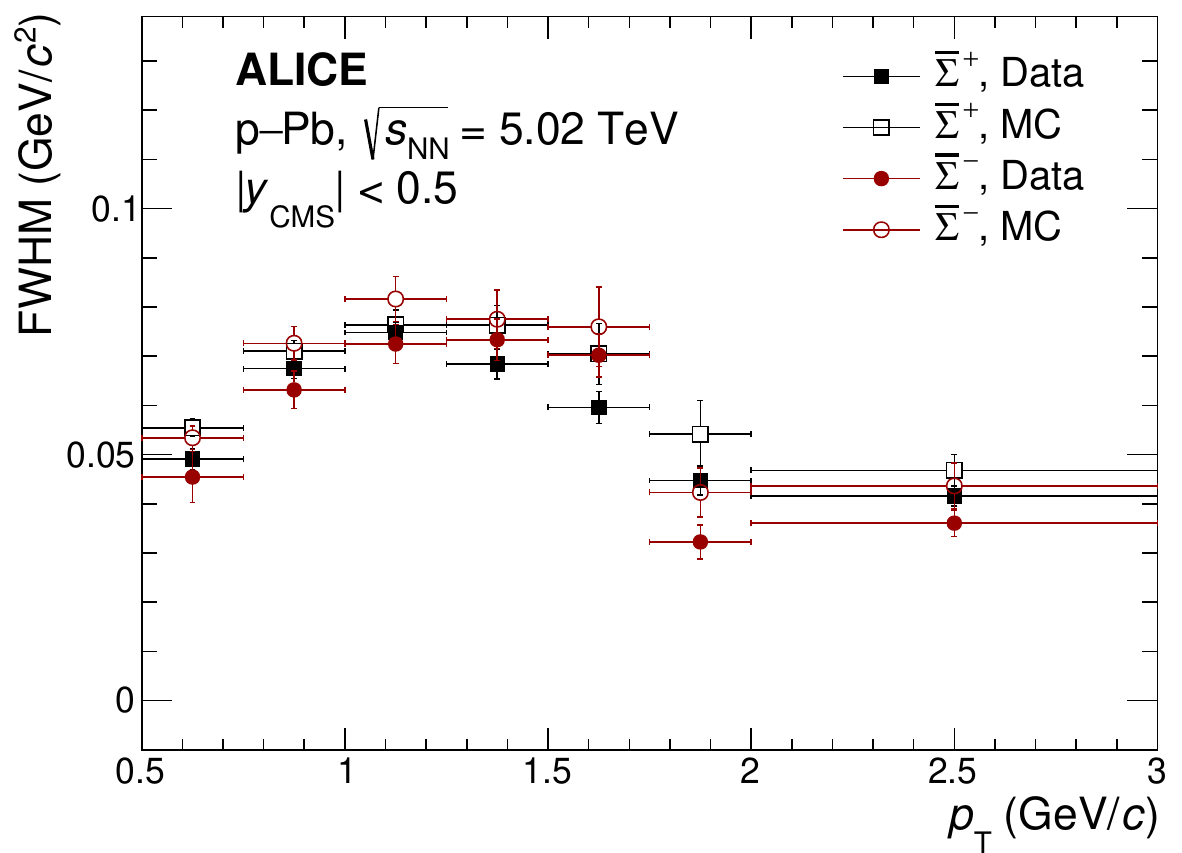}
  \caption{Comparison of peak position (left column) and full width at half maximum, FWHM, (right column) in pp (top row) and p--Pb collisions (bottom row) in data and MC  simulations. The PYTHIA~8~\cite{pythia} and DPMJET 3.0-5~\cite{dpmjet:2000he} event generators were used for Monte-Carlo simulations. The PDG mass values of $\overline{\Sigma}^{+}$ and $\overline{\Sigma}^{-}$~\cite{ParticleDataGroup:2024cfk} are shown with short and long dashed lines, respectively.}
  \label{fig.peakPositionWidth}
\end{figure}

The signal after background subtraction is compared to MC simulations, where $\nbar\pi^\pm$ pairs from the same 
$\psigpmbar$ are selected ("Signal, MC" distribution in Fig.~\ref{fig:MpiN}).
The width of the $\psigpmbar$ peak in the $\nbar\pi^\pm$ invariant mass distribution is defined mostly by the timing resolution of the PHOS. The specific non-Gaussian shape of the peak is related to the time in the denominator in the Eq. (\ref{eq:momentum}). 
Using invariant mass distributions of $\nbar\pi^\pm$ pairs with common parent $\psigpmbar$ both in pp and \ppb MC simulations with the DPMJET 3.0-5~\cite{dpmjet:2000he} event generator, it is found that a sufficiently good description of the shape of the peak can be obtained as a combination of two exponential distributions with independent decay widths on the left and right sides of the peak. 
A comparison of Signal distributions in data and MC is used to check the description of the PHOS response.
As the GEANT 3~\cite{Brun:1119728} description of the \nbar interaction in a calorimeter is rather schematic, a parameterization of the experimentally measured PHOS time response to antibaryons is used in Monte-Carlo. 
To this purpose, clusters matched with antiproton tracks identified in the central tracking system are selected and the difference between the reconstructed time and expected time of antiproton arrival in the front surface of the PHOS which accounts for a track curvature is calculated. 
The width of this time distribution is defined by PHOS electronics timing resolution and by fluctuations in the depth of the hadronic shower in the calorimeter. 

The timing resolution of the PHOS decreases with the cluster energy approximately as $1/E_\mathrm{clu}$ and is estimated to be $\sim 15$~ns at $E_\mathrm{clu}=0.6$~GeV and $\sim7.5$~ns at $E_\mathrm{clu}=1.2$~GeV. 
The time distributions calculated for antiprotons for each value of $E_\mathrm{clu}$ in the range used in the analysis are parameterized and applied to calculate time response in the MC simulations. 
The quantitative comparison of peak positions and FWHM of $\psigpmbar$ signal peaks in data and MC is presented in Fig.~\ref{fig.peakPositionWidth}. A good agreement between data and MC distributions within statistical uncertainties in the full $\pt$ range is found, both in \pp and \ppb collisions, giving additional validation of the proper description of PHOS response.

To calculate the raw yield of $\psigpmbar$, a bin content counting procedure is used. The limits of the bin counting are chosen as a compromise between integrating the larger part of the signal and picking up statistical fluctuations from the background at the long tails of the peak.
As a result, integration ranges corresponding to the full width at 1/3 maximum (or 1/4 maximum for systematics studies) are applied. These ranges are shown in Fig.~\ref{fig:MpiN} with vertical dashed and solid lines for 1/3 and or 1/4 maximum height, respectively.

\begin{figure}[h]
  \centering
  \includegraphics[width = 0.48\linewidth]{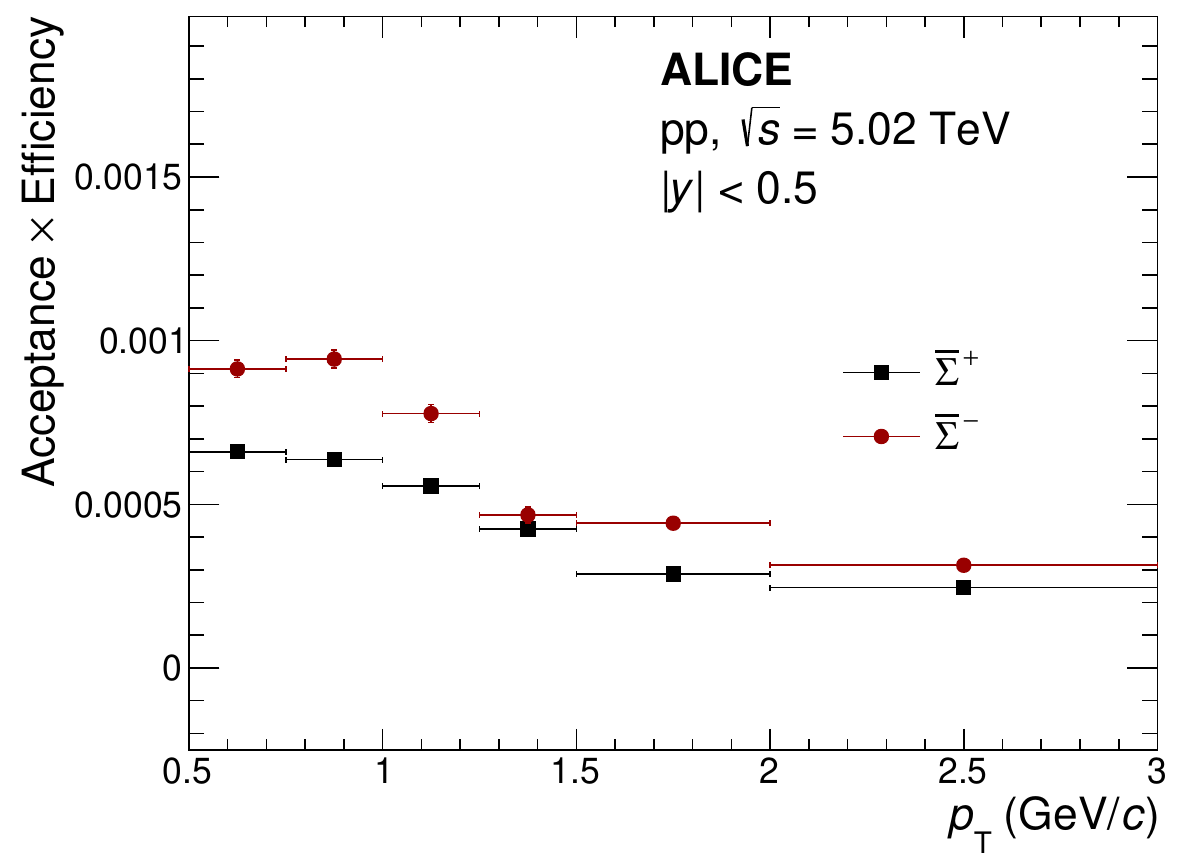}
  \hfill
  \includegraphics[width = 0.48\linewidth]{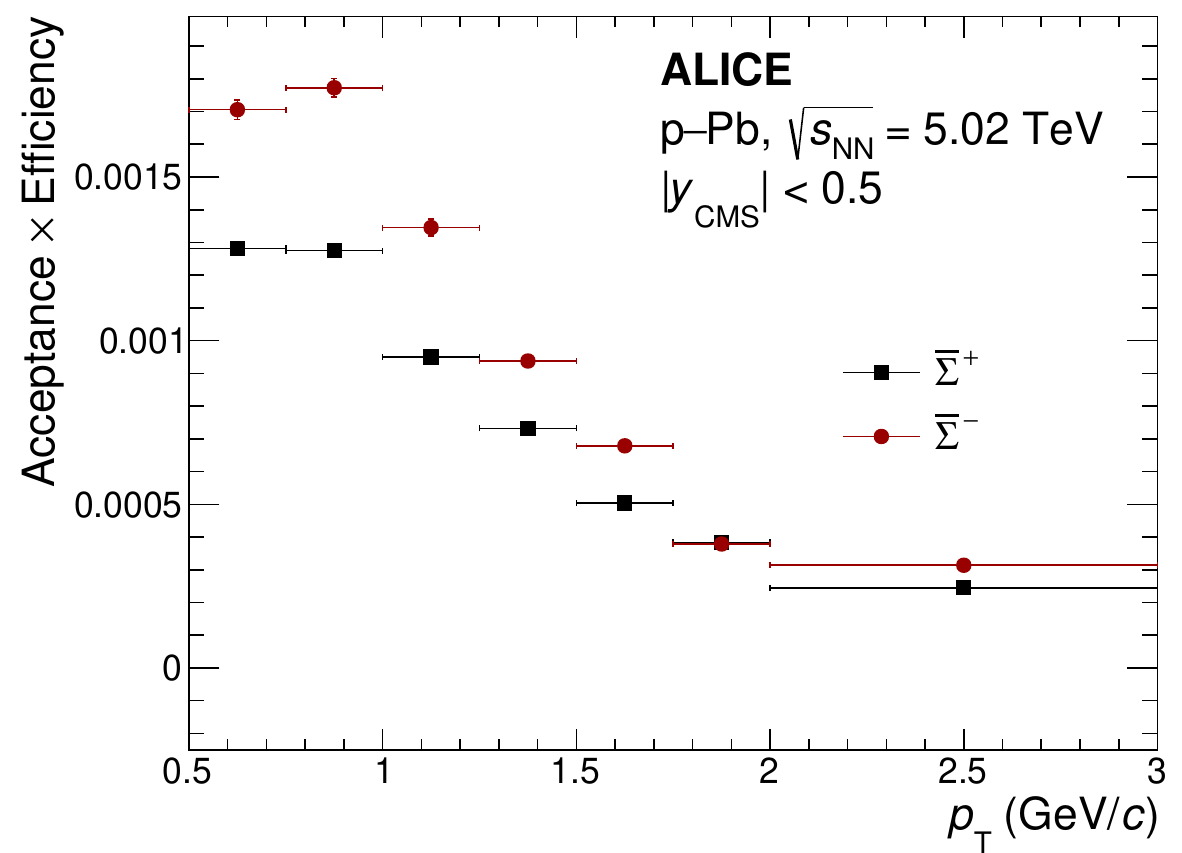}
  \caption{Reconstruction efficiency for $\overline{\Sigma}^{\pm}$ in pp collisions calculated using PYTHIA~8~\cite{pythia} (left) and $\overline{\Sigma}^{\pm}$ in p--Pb collisions (right) calculated using DPMJET 3.0-5~\cite{dpmjet:2000he}.}
  \label{fig:effRec}
\end{figure}

Reconstruction efficiency (see Fig.~\ref{fig:effRec}) is defined as the ratio of the number of $\nbar\pi^\pm$ pairs that pass all selections and originate from the same
$\psigpmbar$ to the generated spectrum of $\psigpmbar$ hyperons.
The reconstruction efficiencies of $\psigpmbar$ were estimated within the AliRoot framework 
using the PYTHIA~8.2.10~\cite{pythia} event generator with Monash 2013 tune for pp collisions and DPMJET 3.0-5~\cite{dpmjet:2000he} for \ppb collisions. 
As both generators do not reproduce the shape of the $\psigpmbar$ $\pt$ spectrum and as the large PHOS timing resolution introduces a dependence on this shape, a weighting procedure was implemented to reproduce the final measured spectrum. 
The introduction of the weights modified the efficiencies and thus the final spectra, therefore an iterative procedure was applied which converges to an accuracy better than statistical uncertainties in 2--3 iterations. 
Similarly, the difference in local slopes of the spectra at low \pt in \pp and \ppb collisions results in a considerable change in the efficiencies.

The shape and the absolute value of the efficiencies are related to the PHOS acceptance and the proportion of $\nbar$ for which 
the measured time is larger than the time of flight for photons, and thus which momenta can be reconstructed.
Other \nbar selection criteria and pair topological selections do not affect the shape and only further reduce the reconstruction efficiencies. Because of the different mean lifetimes of \psigpbar and \psigmbar, different selections on PV to SV distance were applied, resulting in some difference in reconstruction efficiencies of  \psigpbar and \psigmbar.

The $\psigpmbar$ invariant yields are calculated by correcting the raw yields for the reconstruction efficiencies and the V0M
trigger efficiency. The trigger efficiency was estimated in previous analyses and found to be $0.964\pm0.031$ for non-single diffractive (NSD) \ppb collisions~\cite{ALICE:2012xs} and $0.757\pm0.019$  for INEL \pp collisions~\cite{ALICE:2012fjm}.

\FloatBarrier
\subsection{Systematic uncertainties}

Systematic uncertainties can be split into several classes: cluster selection, track selection, topological selections, raw yield extraction, trigger selection, and material budget. The summary of relative uncertainties for $\psigpmbar$ yields is presented in Table~\ref{table:Sys}.

The cluster selection class reflects the quality of the description of the PHOS response to \nbar in MC simulations and combines uncertainties related to $\nbar$ reconstruction and identification in the PHOS.
These uncertainties are estimated by comparing the relative difference between the fully corrected spectra calculated with the default set of selections and with selection variations. 
To reduce statistical fluctuations, the relative differences as a function of $\pt$ are parameterized with a smooth function. 
Linear or constant functions provide a good description in all cases.
If several variations are considered, a maximal deviation is used as an uncertainty estimate. 
Shower shape uncertainties include a variation of the minimum number of cells in the cluster in the range 6--8 and of the offset parameter $a_\mathrm{disp}$ of the dispersion selection in the range 3.5--4.5 cm$^2$. 
As dispersion parameters and the number of cells are correlated, they are varied simultaneously.
The cluster neutrality criterion CPV $n_{\sigma}$ is varied in the range 4--11$\sigma$. 
Minimal cluster energy threshold is varied in the range $0.5< E_{\text{clu}}< 0.7$~GeV.
Uncertainties related to the description of the PHOS time response to neutrons in MC simulations are estimated by varying the width of the parameterized time line by 10\%.

\begin{table}[t]
  \caption{Relative systematic uncertainties of $\psigpbar$ (left value) and $\psigmbar$ (right value) yields in several $\pt$ bins in \pp and \ppb collisions.}
  \label{table:Sys}
  \resizebox{\textwidth}{!}{%
    \begin{tabular}{|c|c|c|c|c|c|c|}
    \hline
      Source                          & \multicolumn{6}{c|}{Uncertainty (\%) $\psigpbar$ / $\psigmbar$}                                                                            \\                        \hline
                                      & \multicolumn{3}{c|}{\pp}             & \multicolumn{3}{c|}{\ppb}                                                 \\
      \hline
      $\pt$ (\gevc)         & 0.5--0.75     & 1--1.25       & 1.5--2          & 0.5--0.75    & 1--1.25      & 1.75--2   \\
      \hline
      Shower shape                    & \multicolumn{3}{c|}{11.4/14.8}                 & \multicolumn{3}{c|}{7.2/11.6}           \\  \hline
      CPV $n_{\sigma}$                & \multicolumn{3}{c|}{3.9/2.7}                   & \multicolumn{3}{c|}{2.5/5.1}            \\  \hline
      Minimum $E_{\text{clu}}$        & \multicolumn{3}{c|}{13.6/10.5}                 & \multicolumn{3}{c|}{11.0/13.6}           \\  \hline
      PHOS time response              & 4.9/5.1     & 5.5/5.8     & 9.4/8.9            & 5.5/4.1    & 7.9/6.5    & 11.5/14.6  \\  \hline  
      Track $|\eta_\mathrm{max}|$     & 3.7/1.0     & 1.4/1.0     & 1.3/1.9            & 4.7/4.5    & 3.4/3.9    & 1.4/3.0   \\  \hline
      $\pi^\pm$ $\text{d}E/\text{d}x$ & 0.4/2.2     & 0.4/2.1     & 0.5/2.1            & 1.7/1.7    & 0.8/1.9    & 0.5/2.0   \\  \hline
      Topological selections          & 2.5/8.8     & 1.6/4.9     & 1.3/5.3            & 6.9/18.2   & 5.4/13.4   & 3.2/6.2   \\  \hline
      Raw yield extraction            & 2.0/5.2     & 2.3/1.9     & 1.0/2.1            & 5.2/12.2   & 3.1/10.2   & 3.0/4.8   \\  \hline
      Material budget                 & \multicolumn{6}{c|}{4.5}                                                                           \\  \hline
      ITS-TPC matching efficiency     & \multicolumn{6}{c|}{3.0}               
                                                      \\  \hline
      Total                           & 20.3/22.6   & 20.1/21.0   & 21.4/22.2          & 18.7/29.9  & 18.3/26.9  & 19.4/25.9 \\  \hline
    \end{tabular}
  }
\end{table}

The uncertainties related to the track selections are estimated by a variation of the selection on the maximal track pseudorapidity $|\eta_\mathrm{max}|<0.7$ to $|\eta_\mathrm{max}|<0.9$, the minimum number of the TPC clusters in the range 50--70, and the maximal difference between the
measured and expected $\text{d}E/\text{d}x$ values in TPC for pions normalized to the resolution from $2.5\sigma$ to  $3.5\sigma$. 
The uncertainties on the topological selections are estimated by a variation of the \pt-dependent selections: distance of the closest approach between $\psigpmbar$ daughters $DCA_\mathrm{daugh}=0.10$ cm to 0.12 cm, the distance between the primary and secondary vertex in case of \psigpbar 0.3 to 0.4 cm and for \psigmbar 0.2 to 0.3 cm (estimated at \pt$=0.5$ GeV/$c$). The cosine of the pointing angle is varied from $CPA>0.0$ to $CPA>0.5$. All topological selections are considered to be correlated and are varied simultaneously.

The uncertainties of the raw yield extraction are estimated by a variation of the integration range obtained at 1/3 and 1/4 maximum height, the fitting range 1.07--1.15 to 1.45--1.8 GeV/$c^2$, and a variation of the parameterization of the residual correlation contribution with a first order polynomial or exponential function.

Some uncertainties were estimated in previous analyses: those related to the material budget description in MC simulations~\cite{Abelev:2014ffa,Abelev:2012cn},
parameterization of the antimatter interaction cross section in GEANT 3~\cite{ALICE:2023aod} and ITS-TPC matching efficiency~\cite{Abelev:2014ffa}. These contributions are considered $\pt$-independent.
All contributions to systematic uncertainties are added in quadrature to obtain the total systematic uncertainty.

\section{Results and discussion}
\label{sec:results}

\subsection{\pt-differential spectra}
\begin{figure}[ht]
  \centering
  \includegraphics[width=0.485\textwidth]{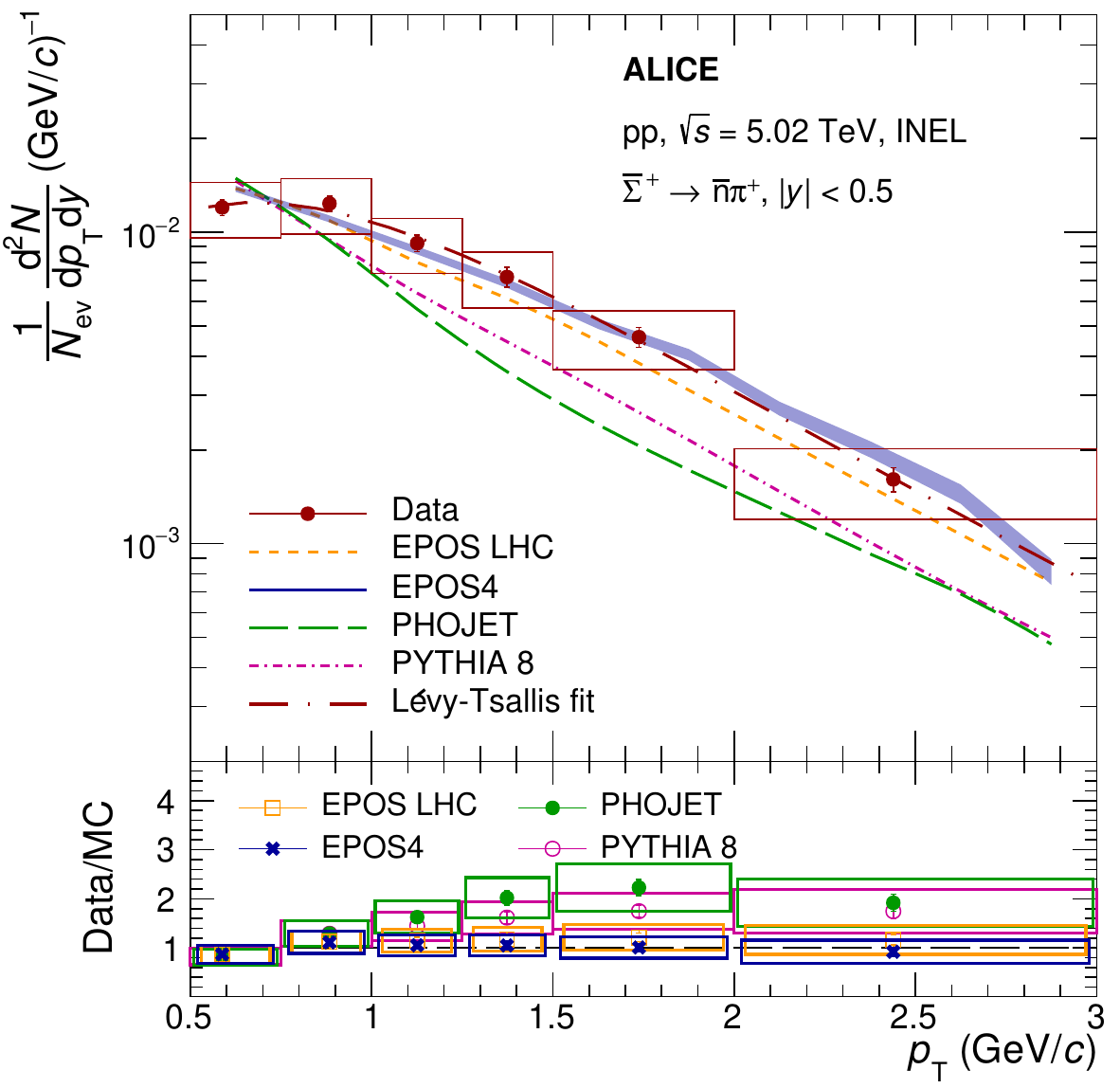}
  \hfill
  \includegraphics[width=0.485\textwidth]{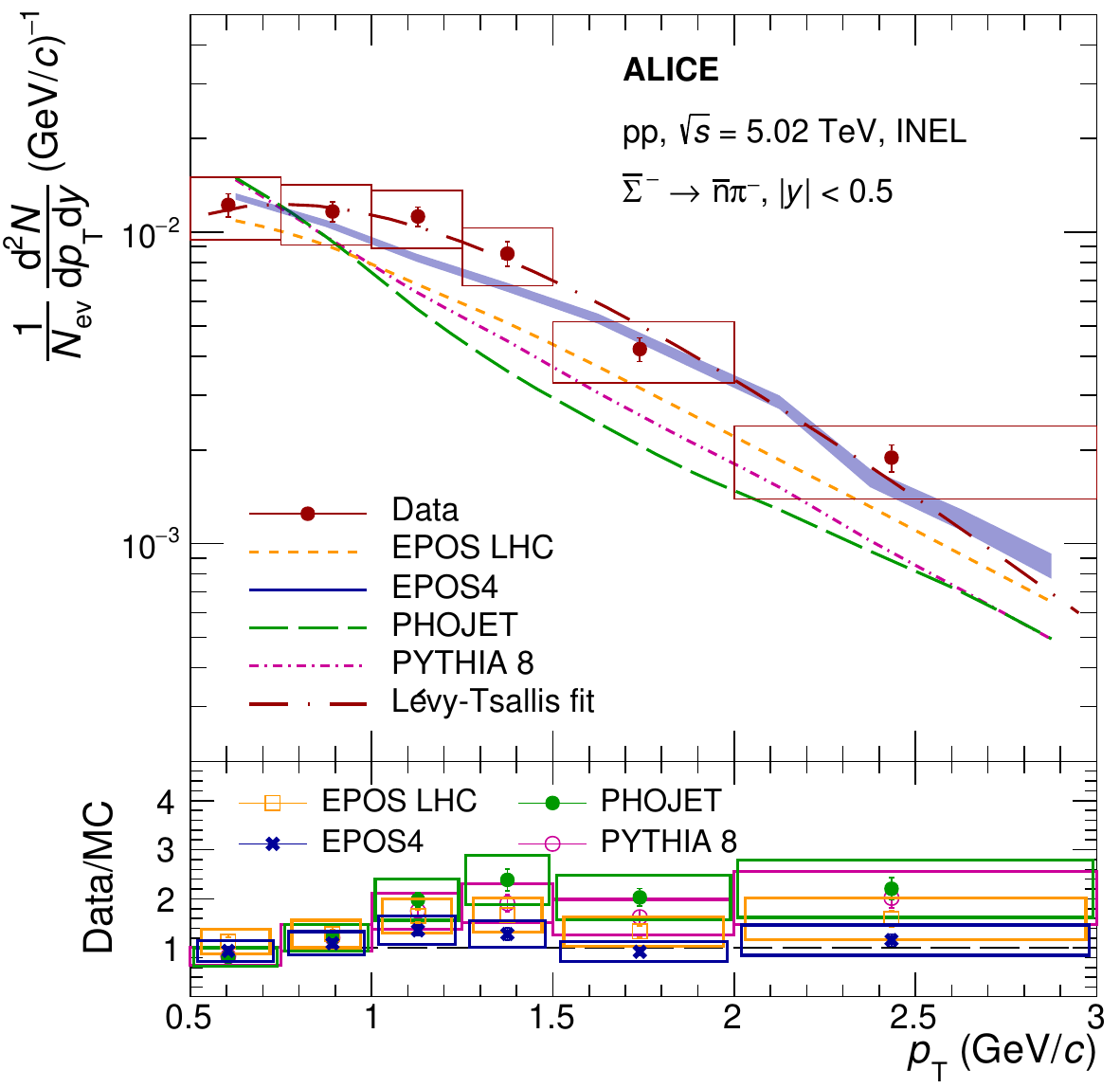}
  \hfill
  \includegraphics[width=0.485\textwidth]{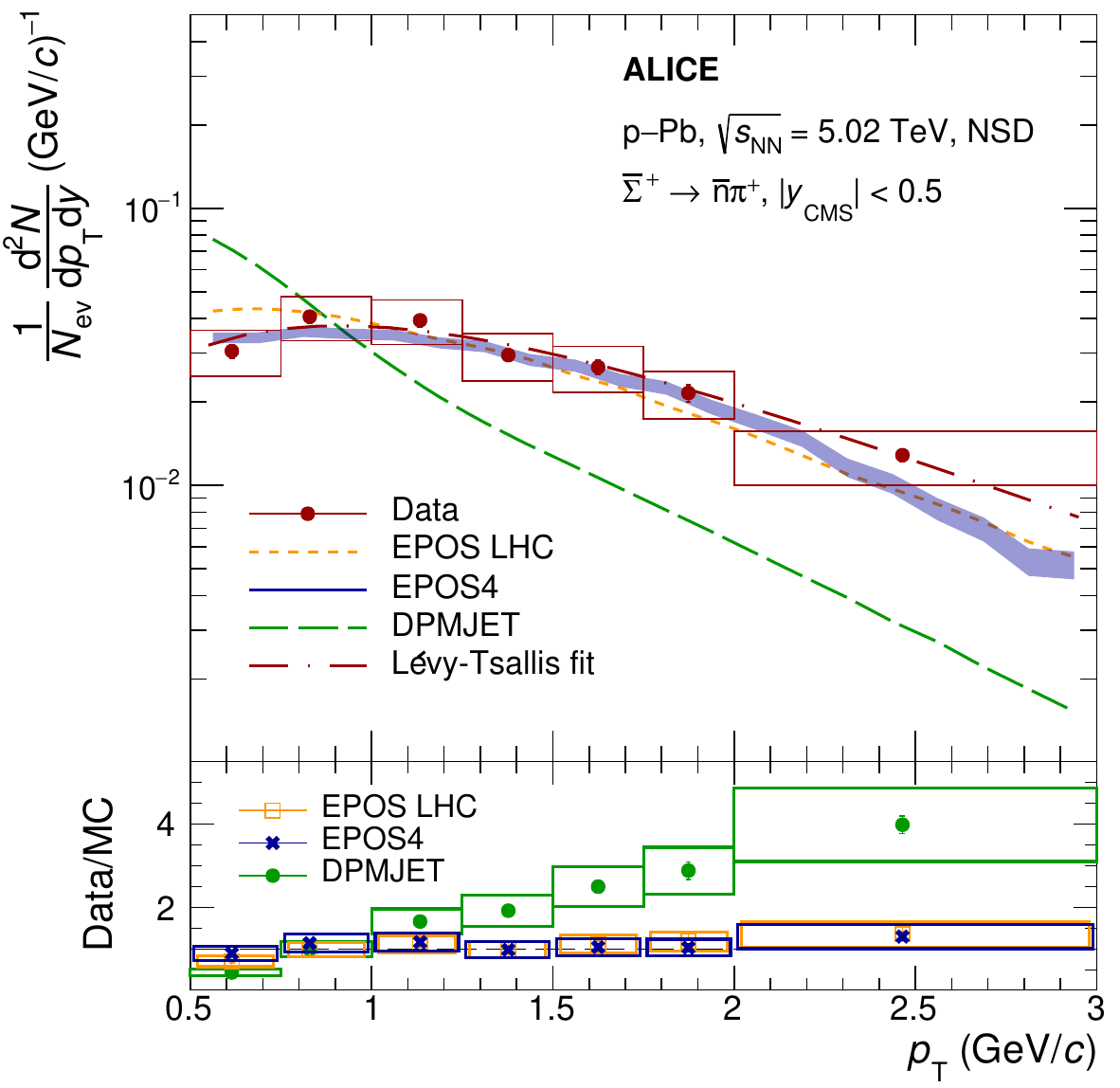}
  \hfill
  \includegraphics[width=0.485\textwidth]{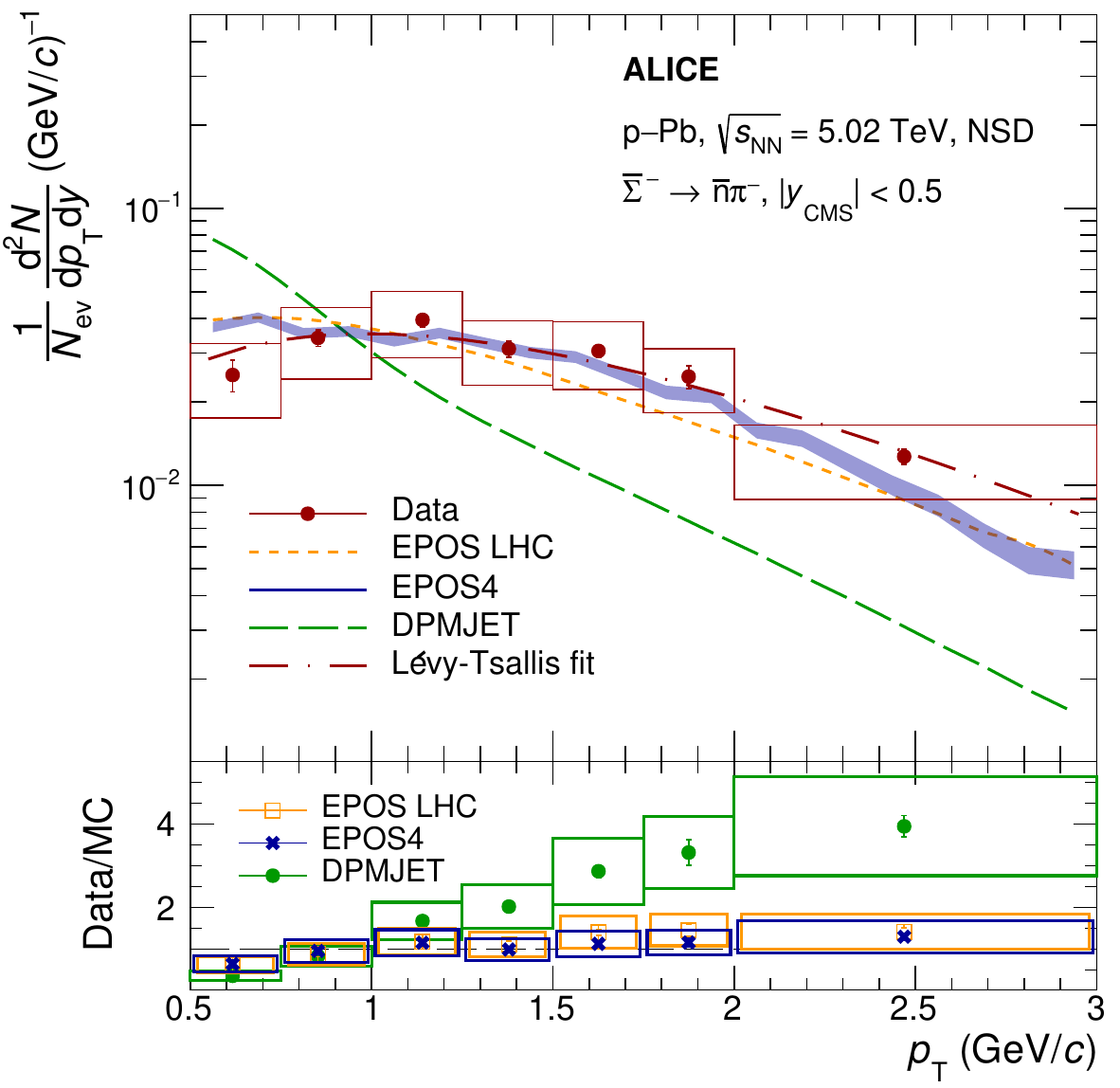}
  \caption{Spectra of $\overline{\Sigma}^{+}$ (left column) and $\overline{\Sigma}^{-}$ (right column) measured in pp (top row) and p--Pb (bottom row) collisions at $\sqrt{s_{\rm{NN}}}~=~5.02~\text{TeV}$ compared to predictions of EPOS LHC and EPOS4~\cite{Pierog:2013ria, Werner:2023zvo}, PYTHIA~8~\cite{pythia}, PHOJET and DPMJET~\cite{Bopp:1998rc} models. Also, Lévy-Tsallis fits to the measured spectra are shown. All models except EPOS4 have negligible statistical uncertainties. The uncertainties of the models are included as statistical ones in the data to MC ratio.}
  \label{fig:Spectra}
\end{figure}

The spectra of $\psigpbar$ and $\psigmbar$ measured in \pp and \ppb collisions are presented in Fig.~\ref{fig:Spectra}. Spectra are extracted in the \pt range of $ 0.5 < \pt < 3$ \gevc.
To account for finite bin widths, point positions are calculated according to prescriptions from Ref.~\cite{Lafferty:1994cj}. The transverse momentum value for each bin, denoted as $p_{\rm T}^{Iw}$, was determined using the mean-ordinate method. This rigorous approach requires solving for the abscissa $p_{\rm T,i}^{Iw}$ in a given bin $i$ such that its functional value equals the mean value of the function over that bin's interval:
\begin{equation}
f(p_{\rm T,i}^{Iw}) = \frac{1}{\Delta p_{\rm T,i}}\int_{p_{\rm T,i}^{\text{low}}}^{p_{\rm T,i}^{\text{up}}} f(p_{\rm T})\, \mathrm{d}p_{\rm T}.
\end{equation}
Since the underlying function $f(p_{\rm T})$ is generally unknown, it was approximated by fitting experimental data with Lévy-Tsallis function to perform this computation.
Measured spectra are compared to predictions of several models.

The best description of the $\pt$ spectra in \pp collisions is provided by the EPOS model (EPOS LHC and EPOS4 versions)~\cite{Pierog:2013ria, Werner:2023zvo} which reproduces measured yields within uncertainties. The EPOS model includes the description of collective effects and combines the parallel multiple scattering scenario (needed in connection with collective effects) and the factorization mechanism. The EPOS4 model was evaluated with all the hydro options switched on ("core full", "hydro hlle", "eos x3ff") and the hadronic re-scatterings simulated with UrQMD ("hacas full").
The PYTHIA  8.2.10 with Monash 2013 tune~\cite{pythia} approximately reproduces the yields at low $\pt$, but predicts 2-2.5 times smaller yields for both hyperons at $\pt>1$ \gevc.
The PHOJET (DPMJET version 3.0-5)~\cite{Bopp:1998rc} model does not reproduce the shape of the spectrum: it approximately agrees with the measured spectrum at low $\pt$, but predicts approximately 3 times smaller yield at high $\pt>1$ \gevc.

In \ppb collisions, the best description of the data is provided by the EPOS LHC and EPOS4 models.
The DPMJET~\cite{dpmjet:2000he} model does not describe the data, predicting considerably higher yield at low \pt and a factor $\sim 4$ smaller yield at high \pt. As the DPMJET model is based on the PHOJET description of nucleon--nucleon collisions scaled accordingly to the Glauber model, the deviation at high $\pt$ inherited from the PHOJET model is expected. The disagreement in the shape may point to the importance of effects like multiparton interactions implemented in the EPOS model, or to the participation of $\Sigma$ hyperons in radial collective flow developed in \ppb collisions.

\begin{figure}[ht]
  \centering
  \includegraphics[width=0.485\textwidth]{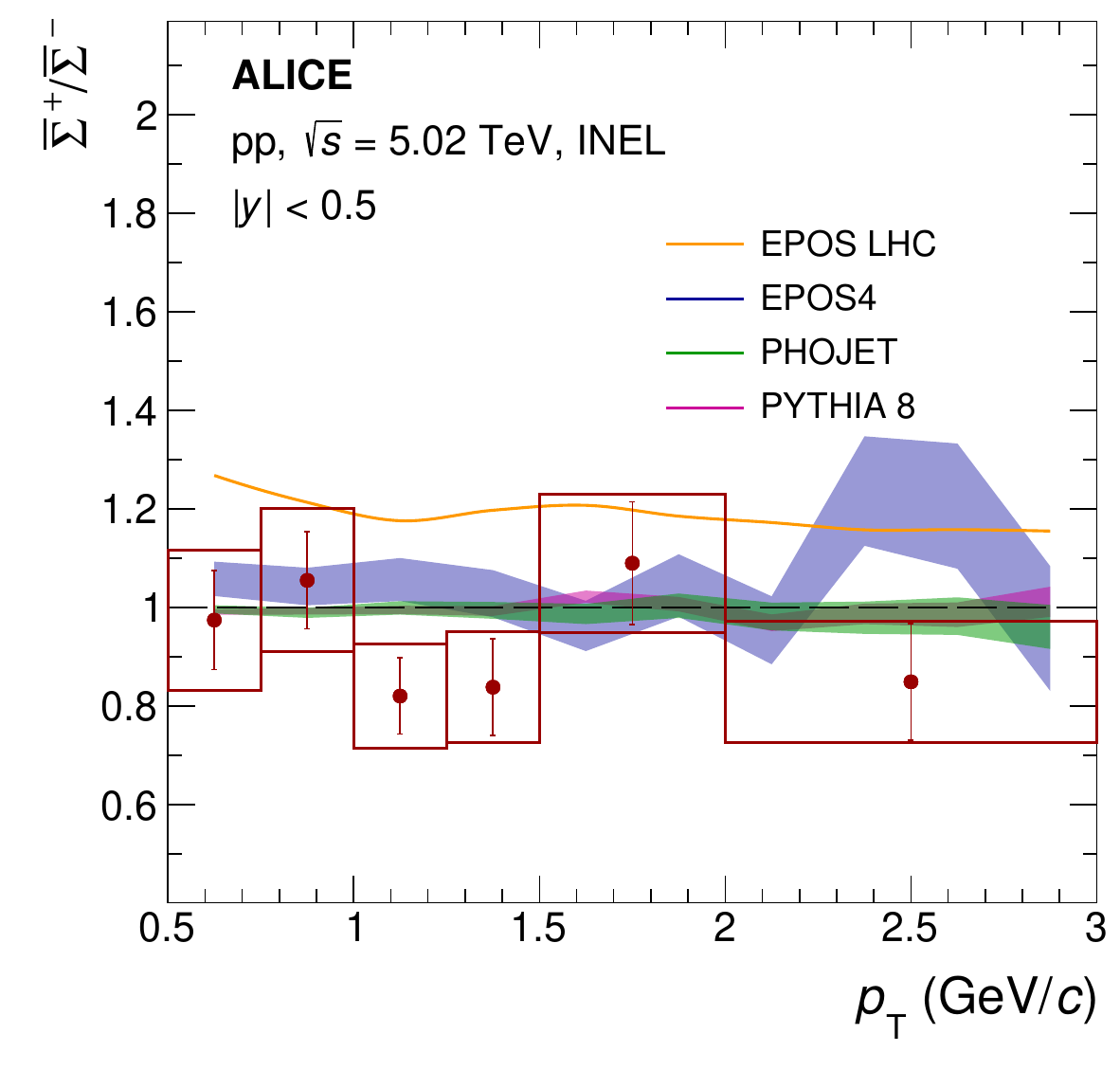}
  \hfill
  \includegraphics[width=0.485\textwidth]{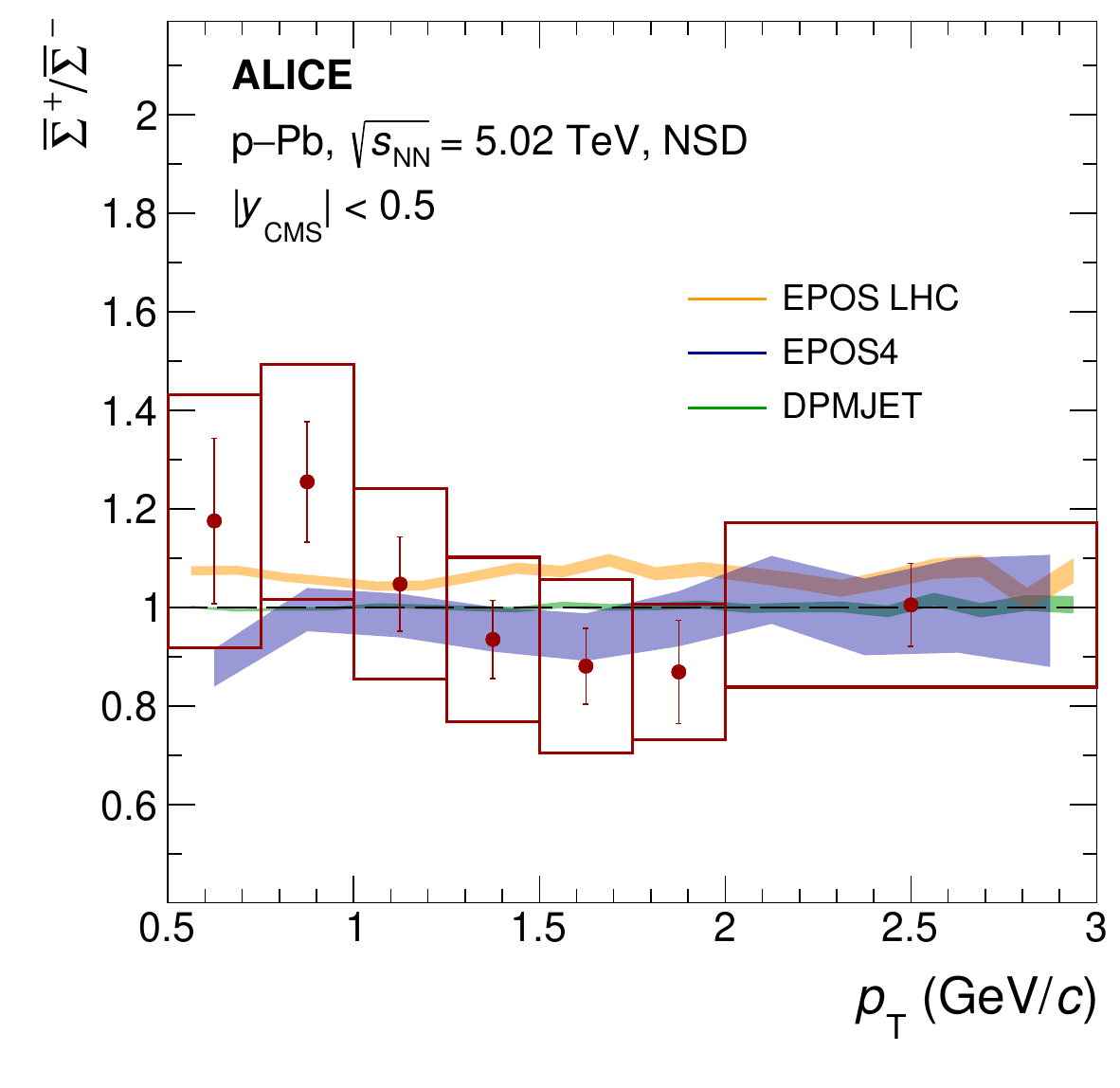}
  \caption{Ratio of $\overline{\Sigma}^{+}$ to $\overline{\Sigma}^{-}$ spectra in pp and p--Pb collisions compared to predictions of EPOS LHC and EPOS4~\cite{Pierog:2013ria, Werner:2023zvo}, PYTHIA~8~\cite{pythia}, PHOJET and DPMJET~\cite{Bopp:1998rc} models.}
  \label{fig:Ratio}
\end{figure}

The ratios of $\pt$ spectra of $\psigpbar$ to $\psigmbar$ measured in \pp and \ppb collisions are presented in Fig.~\ref{fig:Ratio}. 
Some systematic uncertainties related to cluster selections, the PHOS time response parameterization, material budget, etc. cancel out partially or completely in this ratio. 
Because of the quark content of $\psigpbar$ and $\psigmbar$, these ratios are twice as sensitive to the possible deviations from isospin symmetry than, e.g., the n/p ratios.
The measured ratios are consistent with unity, both in \pp and \ppb collisions. The DPMJET, PYTHIA~8 and EPOS4 models predict that these ratios are very close to unity and agree with the data within uncertainties, while the EPOS LHC model predicts some excess of $\psigpbar$ yield compared to $\psigmbar$, especially in pp collisions. This effect could be related to the specific tune of this version of the model because it is not observed in the EPOS4 model predictions.

\begin{figure}[ht]
  \centering
  \includegraphics[width=0.48\textwidth]{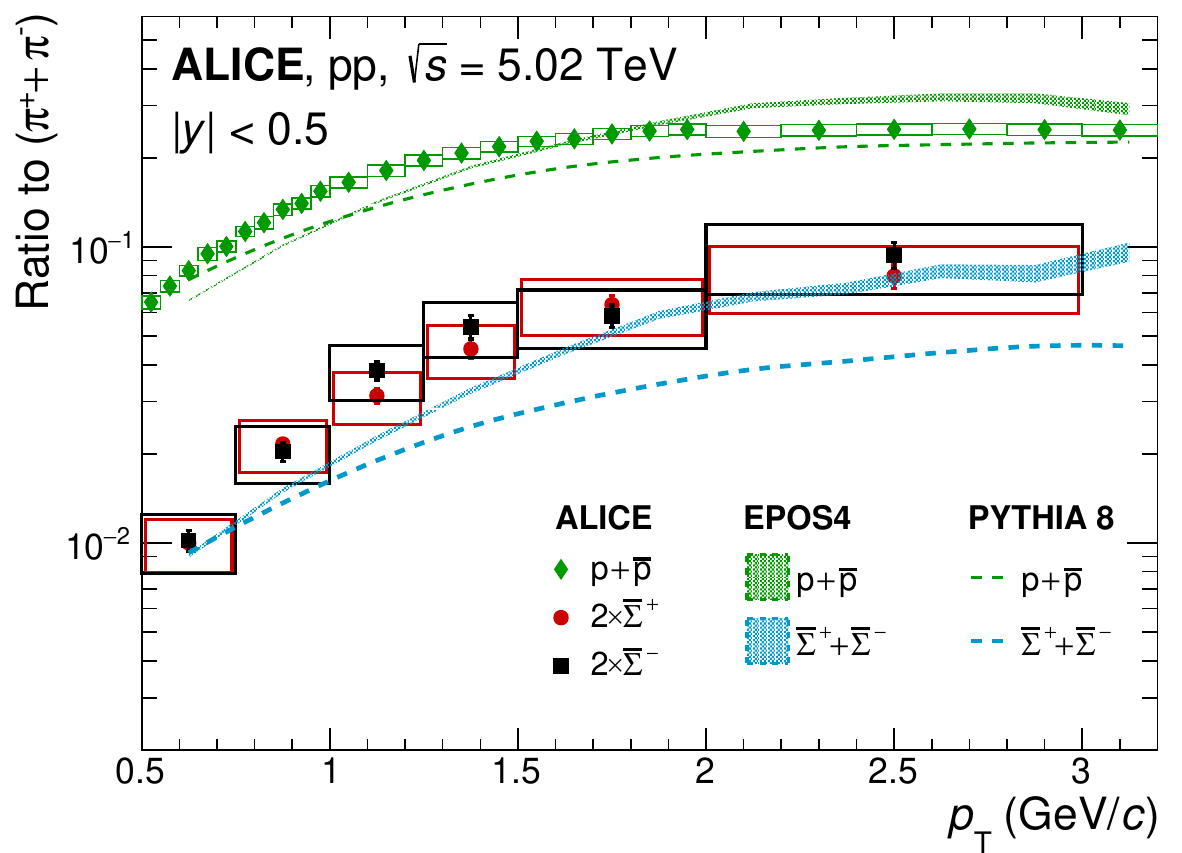}
  \hfill
  \includegraphics[width=0.48\textwidth]{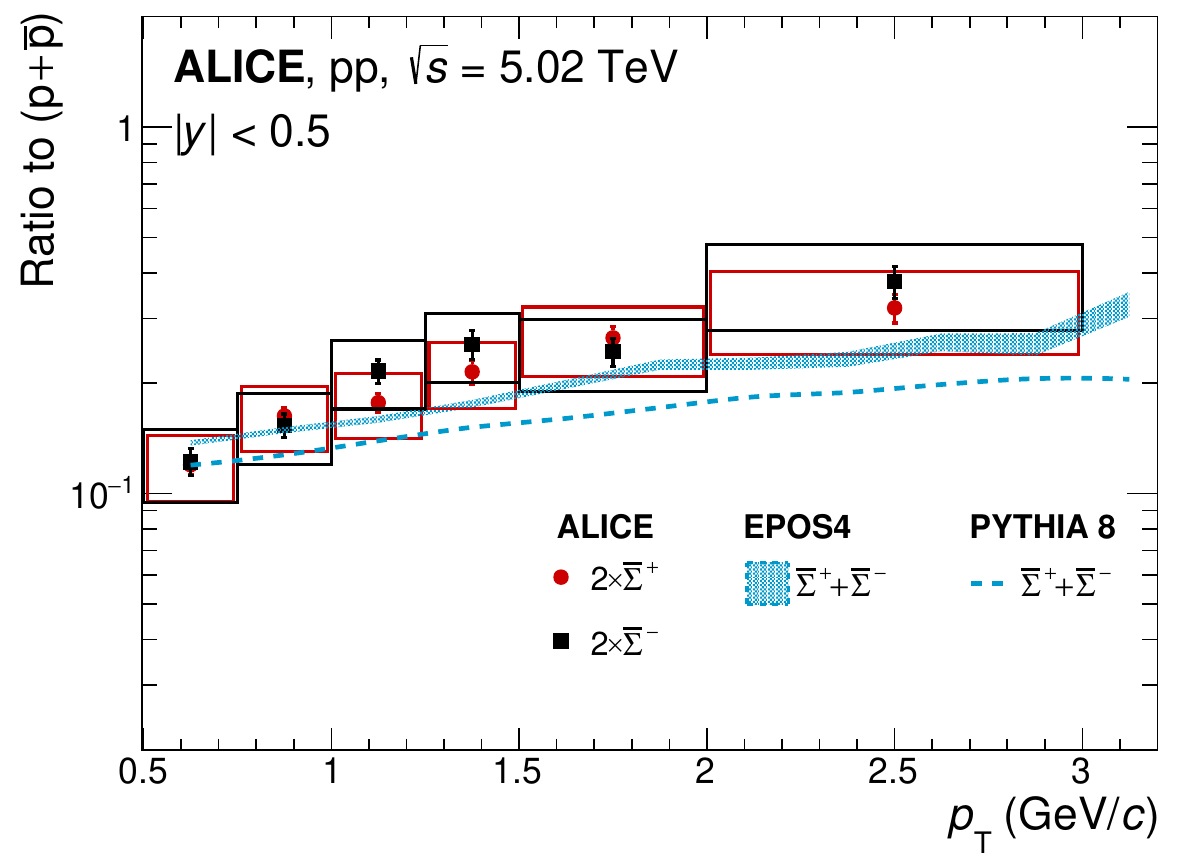}
  \\
  \includegraphics[width=0.48\textwidth]{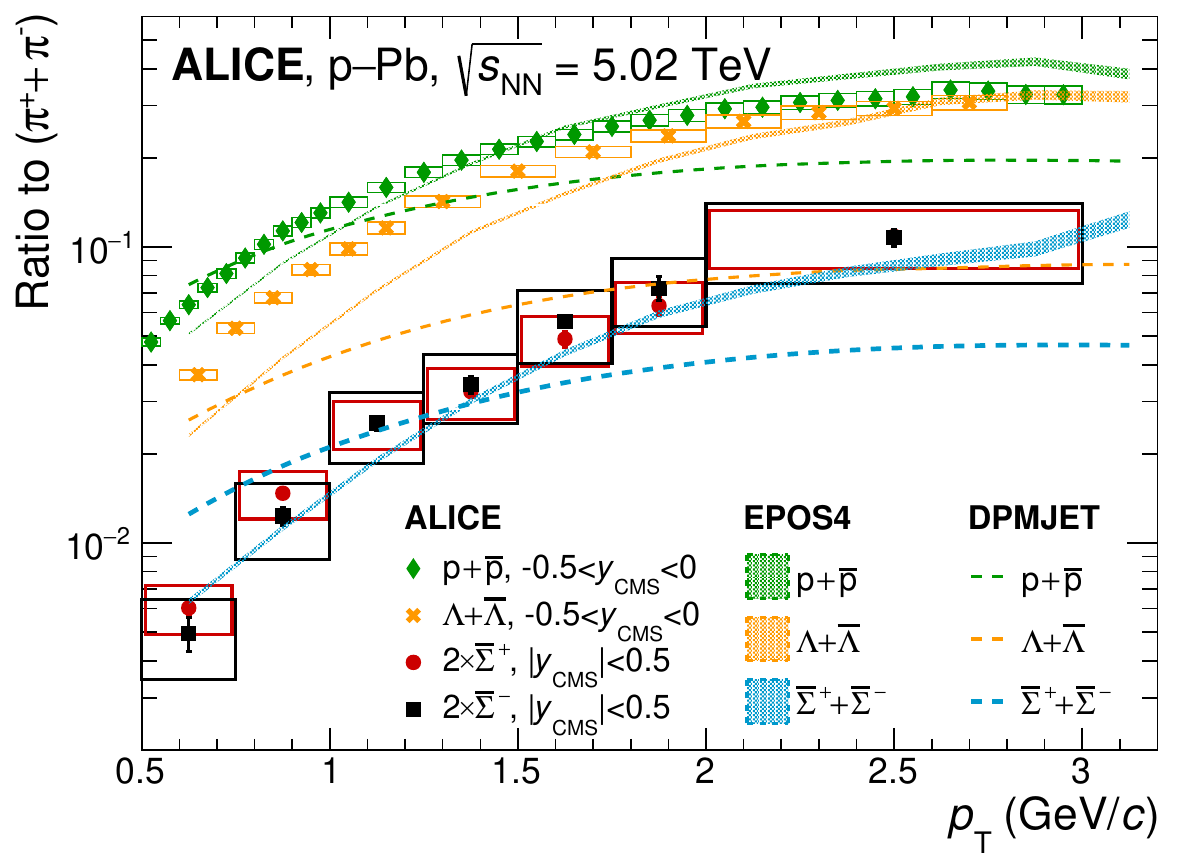}
  \hfill
  \includegraphics[width=0.48\textwidth]{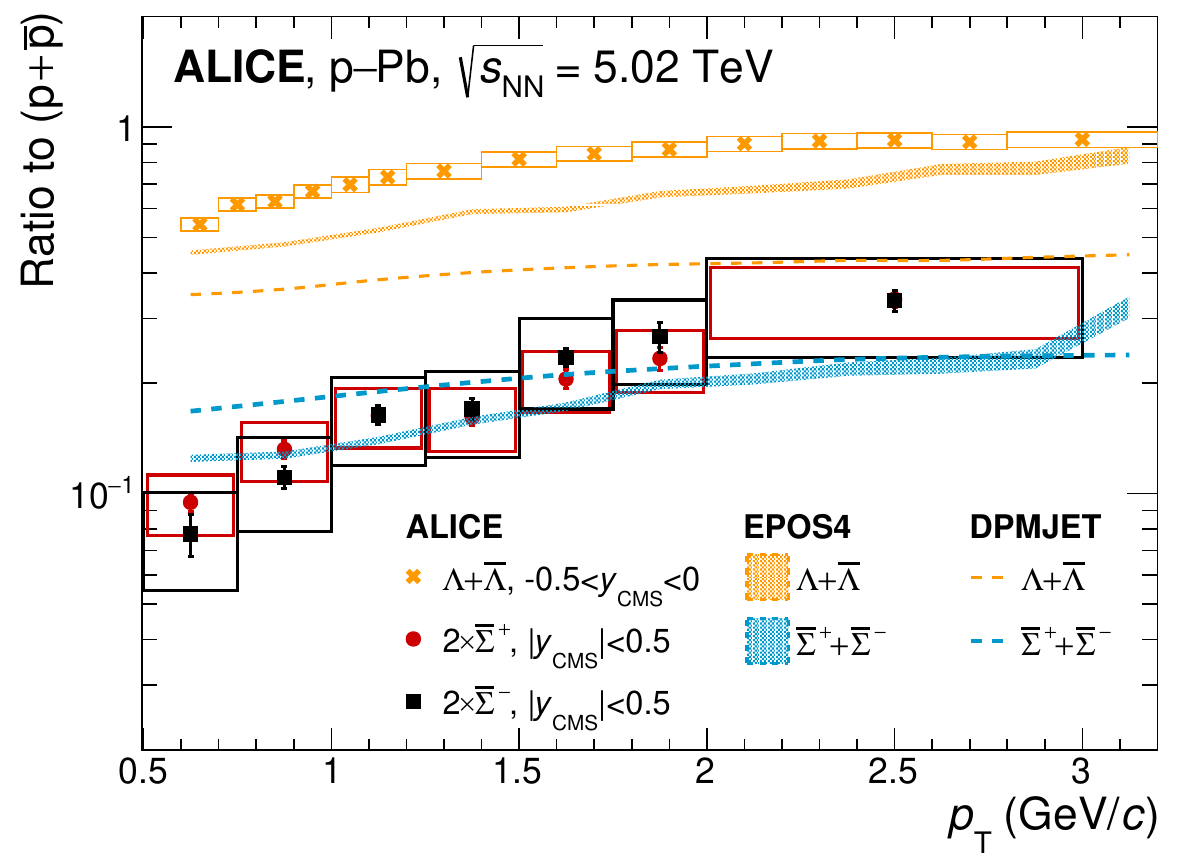}
  \\
  \includegraphics[width=0.48\textwidth]{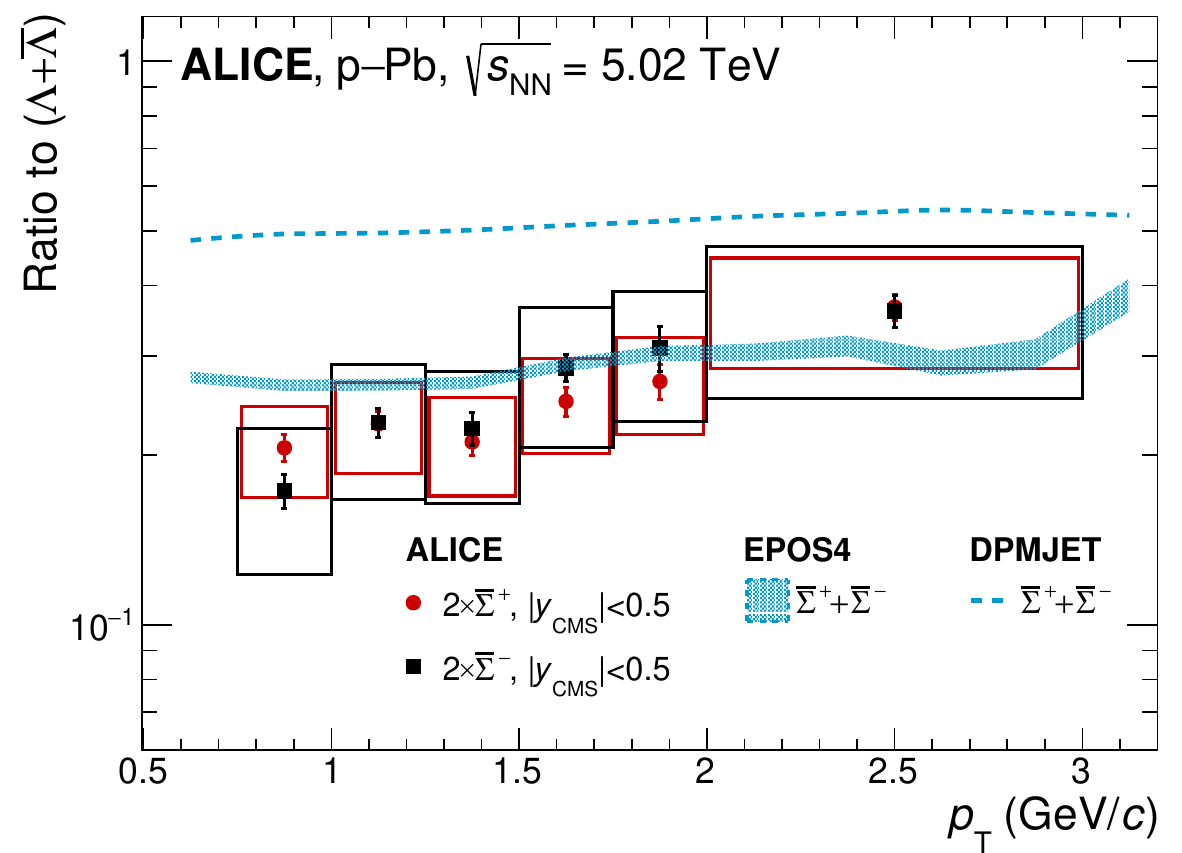}
  \caption{$p_{\rm{T}}$-differential ratios of $\overline{\Sigma}^{-}$, $\overline{\Sigma}^{+}$, $\Lambda$ and protons to pions, protons, and $\Lambda$ (in the case of p--Pb collisions) in pp (top row) and p--Pb (middle and bottom rows) collisions. In the ratios for p--Pb collisions, spectra in the denominator are taken in the rapidity range $-0.5$ < $y_{\rm{CMS}}$ < $0$. Ratios are compared to predictions of EPOS4~\cite{Werner:2023zvo}, DPMJET~\cite{Bopp:1998rc} and PYTHIA~8~\cite{pythia} models.}
  \label{fig:RatioPiPLam}
\end{figure}

Ratios of spectra of different hadron species allow studying in detail different effects like the participation in radial expansion and earlier or later thermodynamical freeze-out for given hadron species, a contribution from coalescence effects, and possible differences in strange and non-strange hard parton fragmentation to mesons or baryons.   
The ratios of spectra of $\psigpmbar$ and different hadrons~\cite{ALICE:2013wgn,ALICE:2019hno} to charged pions, protons, and $\Lambda$ (in the case of \ppb collisions) are compared in Fig.~\ref{fig:RatioPiPLam}.
As the \pt intervals for the \psigpmbar measurements are wider compared to the measured spectra of $\pi$, p, and $\Lambda$, rebinned spectra were used in the denominator of $\psigpmbar/\pi$, $\psigpmbar/$p, and $\psigpmbar/\Lambda$. 
The values for the new bins were obtained from the integral of Lévy-Tsallis fits to the measured spectra. 
The statistical uncertainties of new points are calculated as an integral error from the fit. The systematic uncertainties of new points are calculated by shifting measured points up and down by the systematic uncertainty. The final systematic uncertainty is taken as the maximal deviation of upper and lower values from the central one. 

The p$/\pi$ ratio saturates at high $\pt$ both in \pp and \ppb collisions. 
The $\Lambda/\pi$ demonstrates a similar trend in \ppb collisions.
However, the $\psigpmbar/\pi$ ratio keeps rising to 3 \gevc in \ppb and to a lesser extent in pp collisions. This may point to the participation of $\psigpmbar$ in the radial collective expansion in \ppb collisions. 
Ratios to baryons are less affected by radial expansion because of smaller differences in masses. In addition, a ratio to baryons probes similar production mechanisms at high $\pt$. The $\Lambda/$p ratio shows saturation at high \pt similar to the ratio to pions, but the $\Sigma/$p ratio also shows a rising trend up to the highest measured $\pt$.
EPOS LHC and EPOS4 provide better descriptions of the ratios, especially in \ppb collisions. 
PYTHIA~8 reproduces all ratios at small $\pt$ but predicts smaller baryon proportion at high $\pt$. The DPMJET model does not reproduce  
the observed increase of the ratios with $\pt$
stressing the importance of accounting for collective effects in \ppb collisions.

\FloatBarrier
\subsection{Integrated yield \rm{d}\it{N}/\rm{d}\it{y}}

Reproduction of the integrated hadron yields is one of the basic parameters of dynamical or statistical models. To make a quantitative comparison with model predictions, the integrated yield of \psigpmbar per unit rapidity was estimated. 
As spectra are measured in a limited \pt range, extrapolations are performed: the measured spectra are fitted with a set of functions, including Lévy-Tsallis~\cite{Tsallis:1987eu} (the default one), $p_{\text{T}}$ and $m_{\text{T}}$ exponent distribution, Fermi-Dirac, Boltzmann, and Boltzmann-Gibbs Blast Wave. Fitting was performed using uncertainties calculated as a quadratic sum of statistical and \pt-uncorrelated systematic uncertainties (raw yield extraction), with an additional simultaneous variation of all points with \pt-correlated systematic uncertainties. All functions showed similar and stable results, and the systematic uncertainty was calculated as the RMS for different functions. The fractions of the integrated yields outside the measured region were estimated to be approximately 
16--24\% in the low-$\pt$ region and 1--3\% in the high-$\pt$ region in pp collisions and 
8--14\% in the low-$\pt$ region and 5--12\% in the high-$\pt$ region in the \ppb collisions.
Extracted integrated yields are summarized in Tab.~\ref{tab:yield} and compared to the 
integrated yields calculated by Thermal-FIST (v1.4.2)~\cite{Thermal}, DPMJET, EPOS LHC, EPOS4, and PYTHIA~8 models. All models, including the statistical model Thermal-FIST, reproduce measured yields within uncertainties, both in \pp and \ppb collisions.
Parameters obtained in parameterization of yields with Thermal-FIST model 
are $T = 148.69 \pm 0.03$ MeV, $\mu_{\text{B}}= 0 \pm 0.09$ MeV and 
$T = 147.31 \pm 0.06$ MeV, $\mu_{\text{B}}=0 \pm 0.2$ MeV
for \pp and \ppb collisions, respectively.

\begin{figure}[h]
  \centering
  \includegraphics[width=1\textwidth]{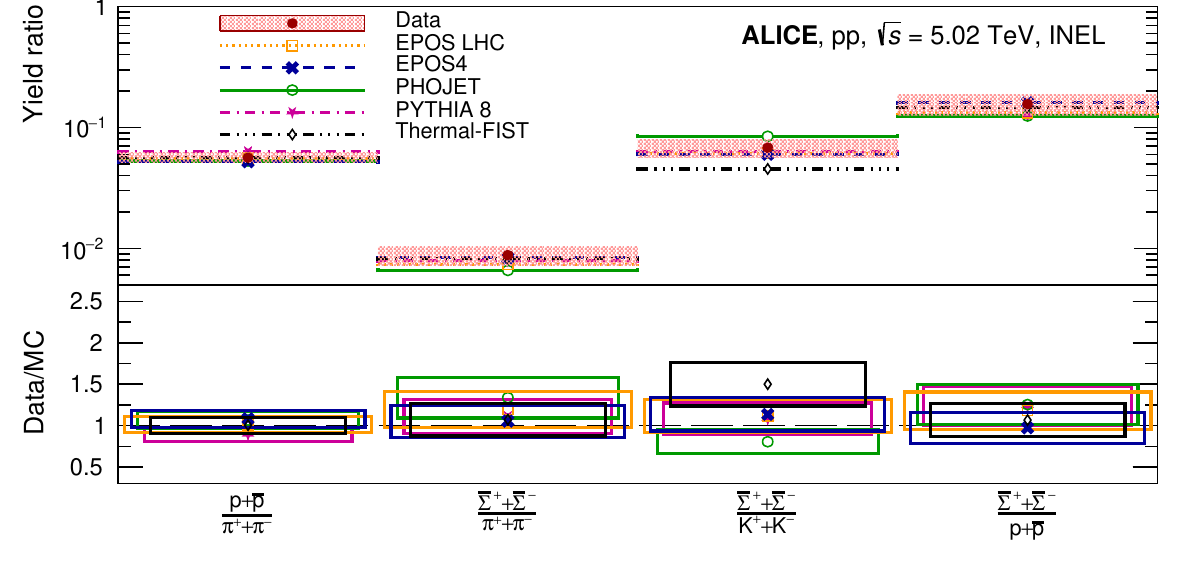}
  \hfill
  \includegraphics[width=1\textwidth]{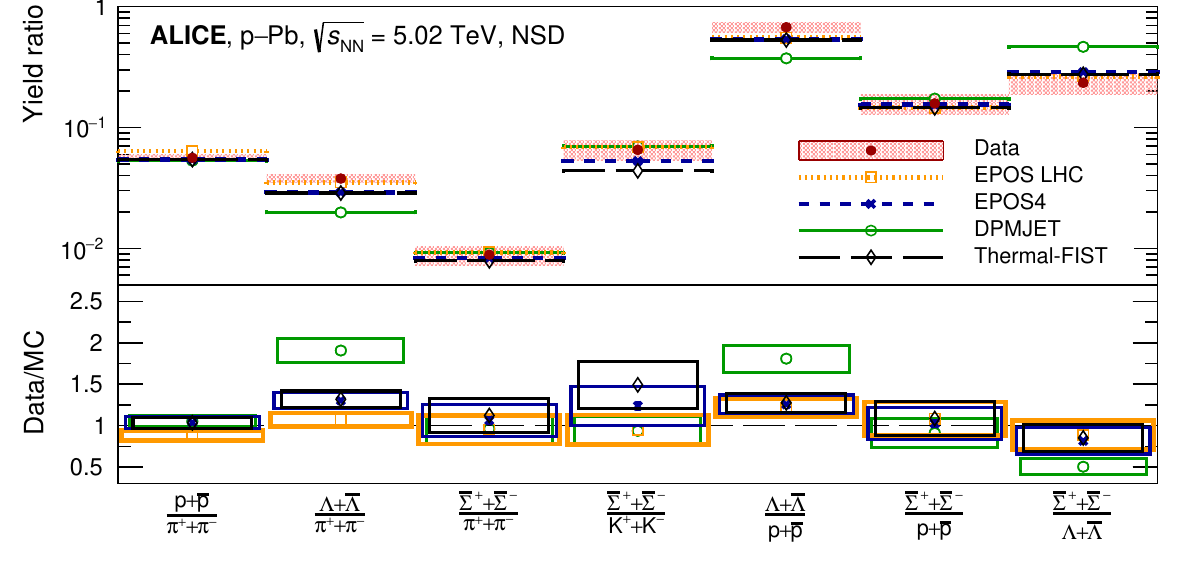}
  \caption{Comparison of ratios of integrated yields of different particle species as shown in Tab.~\ref{tab:ratios} ($\overline{\Sigma}^{-}$ and $\overline{\Sigma}^{+}$ yields are combined). Ratios are compared to predictions of EPOS LHC and EPOS4~\cite{Pierog:2013ria, Werner:2023zvo}, PHOJET and DPMJET~\cite{Bopp:1998rc}, PYTHIA~8~\cite{pythia}, and Thermal-FIST~\cite{Thermal} models.}
  \label{fig:RatioIY}
\end{figure}

Ratios of integrated yields of different hadron species together with model predictions are shown in Fig.~\ref{fig:RatioIY} and summarized in Tab.~\ref{tab:ratios}. Both Thermal-FIST and dynamical model predictions agree with the measurements within uncertainties. It is noteworthy, that the $\Lambda$ baryon yield is difficult to reproduce within dynamical models, and as well the ratio of \psigpmbar to $\Lambda$ generally is not reproduced by dynamical models (with the exception of \psigmbar in the EPOS LHC model), while it is better described by the Thermal-FIST model. On the other hand, the ratio of \psigpmbar to kaons is better reproduced by most dynamical models and underestimated by Thermal-FIST model. These observations might provide insights to the strangeness production mechanisms.

\begin{table}[h]
  \centering
  \caption{Integrated yield of $\psigpmbar$ in \pp and \ppb collisions compared to predictions of several models. Ratios are compared to predictions of EPOS LHC and EPOS4~\cite{Pierog:2013ria, Werner:2023zvo}, PHOJET and DPMJET~\cite{Bopp:1998rc}, PYTHIA~8~\cite{pythia}, and Thermal-FIST~\cite{Thermal} models. For convenience, all yields are multiplied by a factor of $10^3$.} 
  \label{tab:yield}
  \resizebox{\textwidth}{!}{%
  \begin{tabular}{|c|c|c|c|c|c|c|c|}
    \hline
    Specie & $\cfrac{dN}{dy}\pm$ (stat.) $\pm$ (sys.) $\pm$ (extr.) & EPOS4 & EPOS LHC & PHOJET/DPMJET & PYTHIA~8 & Thermal-FIST \\ \hline
                    & \multicolumn{6}{c|}{\pp collisions (INEL), $\times10^3$}                                                                \\ \hline
    $\psigpbar$ & $18.1 \pm 0.5 \pm 4.0 \pm 0.7$ & $21.4 \pm 0.2$ & $18.441 \pm 0.004$ & $16.45 \pm 0.05$ & $17.71 \pm 0.03$ & 16.7 \\ \hline
    $\psigmbar$ & $18.3 \pm 0.7 \pm 4.7 \pm 0.7$ & $20.4 \pm 0.2$ & $14.735 \pm 0.004$ & $16.55 \pm 0.05$ & $17.77 \pm 0.03$  & 17.2 \\ \hline
                    & \multicolumn{6}{c|}{\ppb collisions (NSD), $\times10^3$}                                                                \\ \hline
    $\psigpbar$     & $75 \pm 3 \pm 16 \pm 3$ & $71.9 \pm 0.8$ & $74.1 \pm 0.1$ & $78.39 \pm 0.05$  & --- & 63       \\ \hline
    $\psigmbar$     & $72 \pm 2 \pm 20 \pm 3$ & $73.8 \pm 0.8$ & $69.1 \pm 0.1$ & $78.47 \pm 0.05$  & --- & 65       \\ \hline
  \end{tabular}
  }
\end{table}

\begin{table}[h]
  \centering
  \caption{Ratios of \pt-integrated yields of $\psigmbar$, $\psigpbar$, and $\Lambda$ (in the case of \ppb collisions) to yields of pions, protons, and $\Lambda$ in \pp and \ppb collisions~\cite{ALICE:2013wgn,ALICE:2019hno}. In \ppb collisions, spectra of pions, protons and $\Lambda$ are taken in the rapidity range $-0.5 < y_{\textrm{CMS}} < 0$, other spectra are measured for the rapidity range $|y_{\textrm{CMS}}| < 0.5$. Ratios are compared to predictions of EPOS LHC and EPOS4~\cite{Pierog:2013ria, Werner:2023zvo}, PHOJET and DPMJET~\cite{Bopp:1998rc}, PYTHIA~8~\cite{pythia}, and Thermal-FIST~\cite{Thermal} models. For convenience, all yields are multiplied by a factor of $10^3$.}
  \label{tab:ratios}
  \resizebox{\textwidth}{!}{%
  \begin{tabular}{|c|c|c|c|c|c|c|}
    \hline
    \multicolumn{7}{|c|}{\pp collisions (INEL)}   \\ \hline
    Species & val. $\pm$ (stat.) $\pm$ (sys.) & EPOS4 & EPOS LHC & PHOJET & PYTHIA~8 & Thermal-FIST \\ \hline
    \multicolumn{7}{|c|}{Ratio to ($\pi^+ + \pi^-$) $\times10^3$} \\ \hline
    $p + \bar{p}$ & $56.38 \pm 0.05 \pm 5.0$ & $52.0 \pm 0.1$ & $55.799 \pm 0.004$ & $52.86 \pm 0.05$ & $63.01 \pm 0.03$ & $56.11$   \\ \hline
    $2\psigpbar$ & $8.8 \pm 0.2 \pm 1.9$ & $8.57 \pm 0.08$ & $8.192 \pm 0.002$ & $6.56 \pm 0.02$ & $7.92 \pm 0.01$ & $8.1$         \\ \hline
    $2\psigmbar$ & $8.8 \pm 0.3 \pm 2.3$ & $8.13 \pm 0.08$ & $6.546 \pm 0.002$ & $6.60 \pm 0.02$ & $7.95 \pm 0.01$ & $8.3$       \\ \hline
    \multicolumn{7}{|c|}{Ratio to (K$^+ + $K$^-$) $\times10^3$} \\ \hline
    $2\psigpbar$ & $68 \pm 2 \pm 15$ & $61.5 \pm 0.6$ & $67.91 \pm 0.02$ & $84.3 \pm 0.2$ & $62.8 \pm 0.1$ & $45$         \\ \hline
    $2\psigmbar$ & $68 \pm 2 \pm 18$ & $58.4 \pm 0.6$ & $54.26 \pm 0.01$ & $84.8 \pm 0.2$ & $62.9 \pm 0.1$ & $46$       \\ \hline
    \multicolumn{7}{|c|}{Ratio to (p + $\bar{\textrm{p}}$) $\times10^3$} \\ \hline
    $2\psigpbar$ & $155 \pm 4 \pm 34$ & $165 \pm 2$ & $146.82 \pm 0.04$ & $124.1 \pm 0.4$ & $125.7 \pm 0.2$ & $144$         \\ \hline
    $2\psigmbar$ & $157 \pm 6 \pm 40$ & $156 \pm 2$ & $117.31 \pm 0.03$ & $124.9 \pm 0.4$ & $126.1 \pm 0.2$ & $149$         \\ \hline
    \multicolumn{7}{|c|}{\ppb collisions (NSD)}                                             \\ \hline
    Species & val. $\pm$ (stat.) $\pm$ (sys.) & EPOS4 & EPOS LHC & DPMJET & PYTHIA~8 & Thermal-FIST \\ \hline
    \multicolumn{7}{|c|}{Ratio to ($\pi^+ + \pi^-$) $\times10^3$} \\ \hline
    $p + \bar{p}$ & $56.1 \pm 0.1 \pm 4.6$ & $54.1 \pm 0.2$ & $63.75 \pm 0.03$ & $53.19 \pm 0.01$ & --- & $54.2$         \\ \hline
    $\Lambda + \bar{\Lambda}$ & $37.8 \pm 0.2 \pm 3.7$ & $29.0 \pm 0.1$ & $35.30 \pm 0.02$ & $19.840 \pm 0.006$ & --- & $28.7$        \\ \hline
    $2\psigpbar$ & $9.0 \pm 0.4 \pm 1.9$ & $8.23 \pm 0.09$ & $9.63 \pm 0.02$ & $9.212 \pm 0.006$ & --- & $7.8$         \\ \hline
    $2\psigmbar$ & $8.7 \pm 0.2 \pm 2.5$ & $8.45 \pm 0.09$ & $8.99 \pm 0.01$ & $9.222 \pm 0.006$ & --- & $8.0$         \\ \hline
    \multicolumn{7}{|c|}{Ratio to (K$^+ + $K$^-$) $\times10^3$} \\ \hline
    $2\psigpbar$ & $67 \pm 3 \pm 15$ & $52.1 \pm 0.6$ & $71.3 \pm 0.1$ & $69.71 \pm 0.04$ & --- & $43$         \\ \hline
    $2\psigmbar$ & $64 \pm 2 \pm 19$ & $53.5 \pm 0.6$ & $66.5 \pm 0.1$ & $69.78 \pm 0.04$ & --- & $44$         \\ \hline
    \multicolumn{7}{|c|}{Ratio to (p + $\bar{\textrm{p}}$) $\times10^3$} \\ \hline
    $\Lambda + \bar{\Lambda}$ & $674 \pm 3 \pm 73$ & $537 \pm 3$ & $553.7 \pm 0.4$ & $373.0 \pm 0.1$ & --- & $529$         \\ \hline
    $2\psigpbar$ & $161 \pm 7 \pm 35$ & $152 \pm 2$ & $151.1 \pm 0.2$ & $173.2 \pm 0.1$ & --- & $143$         \\ \hline
    $2\psigmbar$ & $155 \pm 4 \pm 45$ & $156 \pm 2$ & $140.9 \pm 0.2$ & $173.4 \pm 0.1$ & --- & $148$         \\ \hline
    \multicolumn{7}{|c|}{Ratio to ($\Lambda + \bar{\Lambda}$) $\times10^3$} \\ \hline
    $2\psigpbar$ & $239 \pm 10 \pm 54$ & $283 \pm 3$ & $273 \pm 0.5$ & $464.3 \pm 0.3$ & --- & $271$         \\ \hline
    $2\psigmbar$ & $229 \pm 6 \pm 68$ & $291 \pm 3$ & $254.5 \pm 0.4$ & $464.8 \pm 0.3$ & --- & $279$        \\ \hline
  \end{tabular}
  }
\end{table}

\FloatBarrier

\subsection{Nuclear modification factor \RpPb}
\begin{figure}[h]
  \centering
  \includegraphics[width=0.49\textwidth]{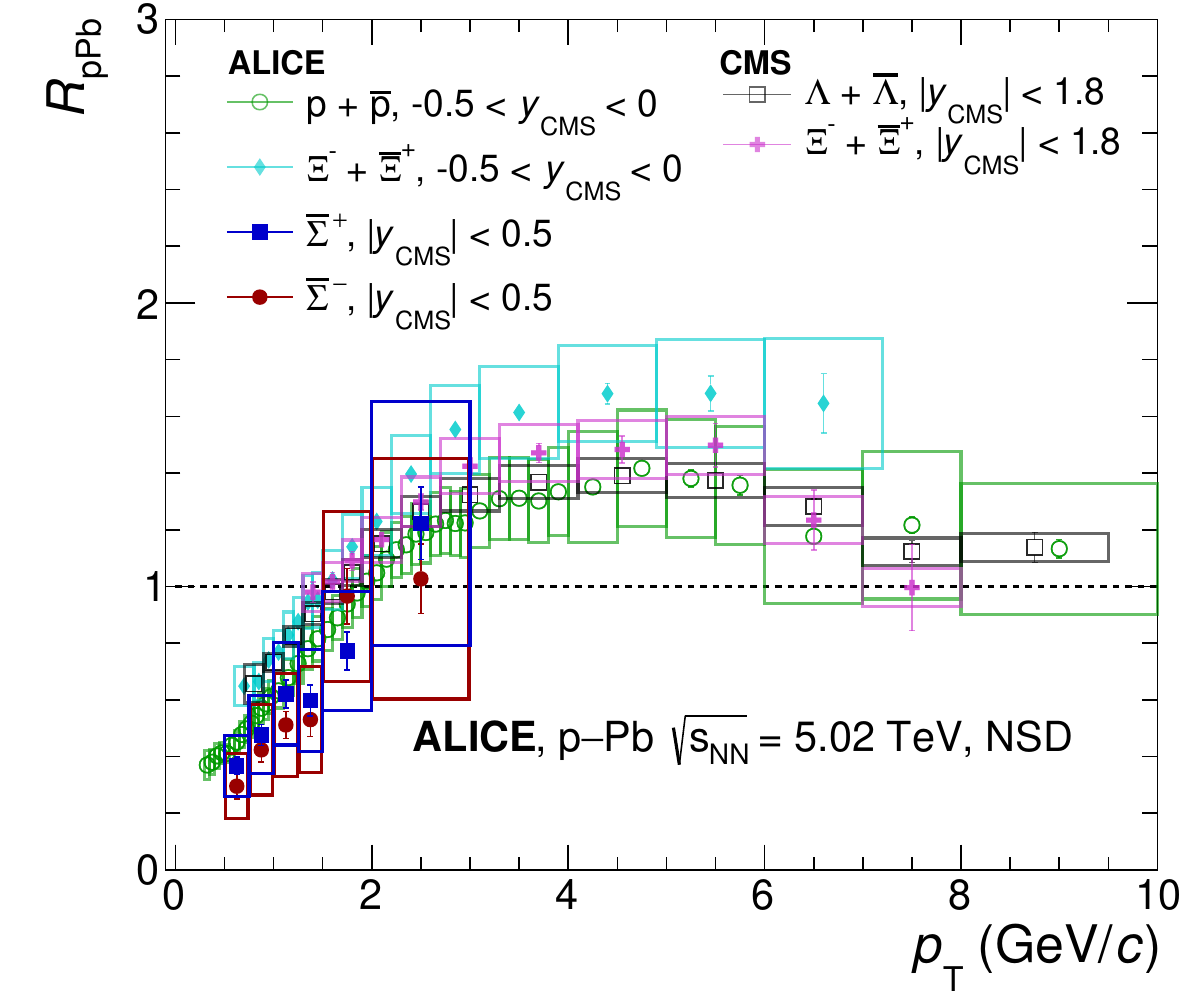}
    \hfill
  \includegraphics[width=0.49\textwidth]{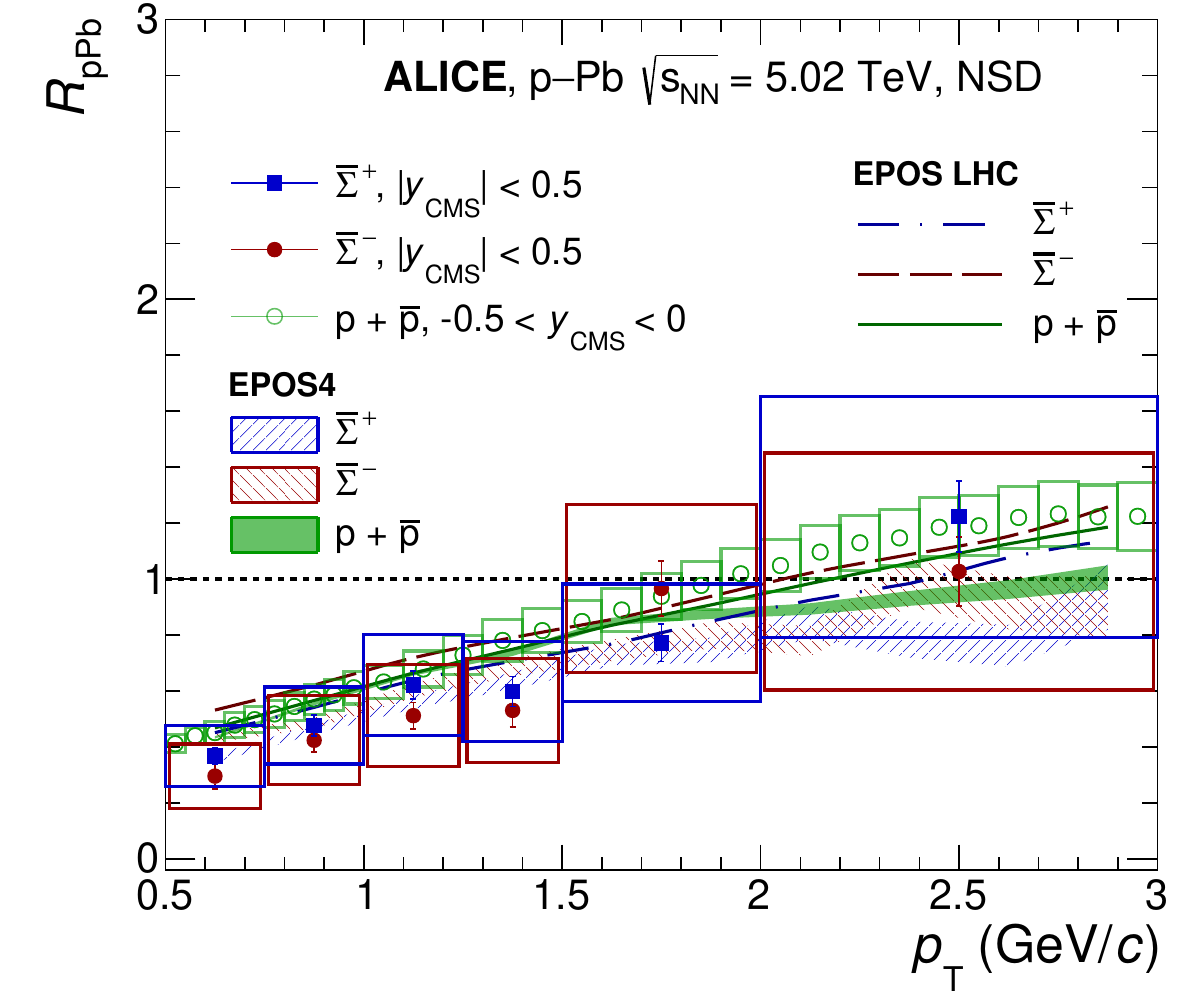}  
    \hfill
  \caption{Nuclear modification factors $R_{\rm{pPb}}$ measured for $\overline{\Sigma}^{+}$ and $\overline{\Sigma}^{-}$ compared to those of protons~\cite{ALICE:2016dei}
and $\Xi$~\cite{CMS:2019isl,ALICE:2016dei} (left) and to predictions of EPOS4 and EPOS LHC models~\cite{Pierog:2013ria} (right).}
  \label{fig:RPpb}
\end{figure}

To compare spectra measured in \pp and \ppb collisions, we construct a nuclear modification factor \RpPb, defined as
\begin{equation}
  \RpPb = \frac{\mathrm{d}N_{\mathrm{pPb}}/d\pt}{\langle N_\mathrm{coll}\rangle\; \mathrm{d}N_{\pp}/d\pt},
\end{equation}
where $\langle N_\mathrm{coll}\rangle = 6.9\pm0.7$ is the number of binary nucleon--nucleon collisions in NSD p--Pb collisions estimated with the Glauber model~\cite{Adam:2014qja}.
The measured $\RpPb$ for $\psigpbar$ and $\psigmbar$ are shown in Fig.~\ref{fig:RPpb} and compared to those of protons~\cite{ALICE:2016dei}, single strange hyperons ($\Lambda$)~\cite{CMS:2019isl}
and hyperons with two strange quarks ($\Xi$)~\cite{CMS:2019isl,ALICE:2016dei}.  
\RpPb for all hadrons agrees within uncertainties, thus no significant dependence on strangeness content was observed. 
The low \pt part deviates from the $\langle N_\mathrm{coll}\rangle$ scaling, which may be caused by the nuclear shadowing of the initial gluon distributions or may show the presence of soft processes, where the number of created soft particles is proportional to the number of wounded nucleons $N_\mathrm{WN}$ in the Glauber model. 
In the right plot of Fig.~\ref{fig:RPpb}, the measured \RpPb of protons, $\psigpbar$, and $\psigmbar$ are compared to predictions of the EPOS LHC and EPOS4 models. 
Both the EPOS LHC and EPOS4 models reproduce the measured \RpPb for $\psigpbar$ and $\psigmbar$ within uncertainties.

\FloatBarrier

\section{Conclusion}
\label{sec:conclusions}

Anti-$\Sigma$ hyperon ($\psigpmbar$) production spectra were measured in
\pp and \ppb collisions at $\sqrt{s_{\mathrm{NN}}}=5.02$ TeV in the transverse momentum range $0.5<\pt<3$ \gevc.
Measurements were performed in $\psigpmbar \rightarrow \nbar\pi^{\pm}$ channel with an \nbar reconstructed in the electromagnetic calorimeter PHOS and $\pi^\pm$ in the central tracking system. 
This is the first measurement to implement the reconstruction of antineutrons with an electromagnetic calorimeter.
The measured spectra of $\psigpmbar$ were compared to 
 PYTHIA~8, EPOS LHC, EPOS4 and PHOJET for \pp and 
DPMJET, EPOS LHC and EPOS4 for \ppb Monte-Carlo models. 
EPOS LHC and EPOS4 reproduce the spectra well both in \pp and \ppb collisions,
while PYTHIA~8 and PHOJET underestimate the measured spectra by a factor of $\sim 3$ at high $\pt$. 
DPMJET predicts a considerably higher yield at low \pt and a yield by a factor of $\sim 4$ times smaller at high \pt. 

The integrated yields of $\psigpmbar$ in both \pp and \ppb collisions were calculated and compared to the Thermal-FIST predictions and to predictions of dynamical models (PYTHIA~8, EPOS LHC, EPOS4, PHOJET, and DPMJET). Both Thermal-FIST and dynamical model predictions describe the integrated yields of $\psigpmbar$ within uncertainties. The ratio of integrated yields of $\psigpmbar$ to the integrated yields of strange and non-strange hadrons were calculated. The current models struggle to describe absolute yields of $\Lambda$ hyperons, and the ratio of $\Sigma$ to $\Lambda$ generally is also overestimated by models, while the ratio to kaons is described better. On the other hand, the Thermal-FIST prediction of $\Sigma$ to $\Lambda$ ratio is consistent with the data. This may provide important insight into the particle production mechanism in high-multiplicity pp and \ppb events. 

The nuclear modification factor $R_\mathrm{pPb}$ was constructed for $\psigpmbar$ and compared to those of the protons, $\Lambda$, 
and $\Xi$ hyperons. \RpPb of $\psigpmbar$ coincide with \RpPb of other baryons within uncertainties. Both the EPOS LHC and EPOS4 models properly reproduce the measured \RpPb of $\psigpmbar$.

\newenvironment{acknowledgement}{\relax}{\relax}
\begin{acknowledgement}
\section*{Acknowledgements}

The ALICE Collaboration would like to thank all its engineers and technicians for their invaluable contributions to the construction of the experiment and the CERN accelerator teams for the outstanding performance of the LHC complex.
The ALICE Collaboration gratefully acknowledges the resources and support provided by all Grid centres and the Worldwide LHC Computing Grid (WLCG) collaboration.
The ALICE Collaboration acknowledges the following funding agencies for their support in building and running the ALICE detector:
A. I. Alikhanyan National Science Laboratory (Yerevan Physics Institute) Foundation (ANSL), State Committee of Science and World Federation of Scientists (WFS), Armenia;
Austrian Academy of Sciences, Austrian Science Fund (FWF): [M 2467-N36] and Nationalstiftung f\"{u}r Forschung, Technologie und Entwicklung, Austria;
Ministry of Communications and High Technologies, National Nuclear Research Center, Azerbaijan;
Rede Nacional de Física de Altas Energias (Renafae), Financiadora de Estudos e Projetos (Finep), Funda\c{c}\~{a}o de Amparo \`{a} Pesquisa do Estado de S\~{a}o Paulo (FAPESP) and The Sao Paulo Research Foundation  (FAPESP), Brazil;
Bulgarian Ministry of Education and Science, within the National Roadmap for Research Infrastructures 2020-2027 (object CERN), Bulgaria;
Ministry of Education of China (MOEC) , Ministry of Science \& Technology of China (MSTC) and National Natural Science Foundation of China (NSFC), China;
Ministry of Science and Education and Croatian Science Foundation, Croatia;
Centro de Aplicaciones Tecnol\'{o}gicas y Desarrollo Nuclear (CEADEN), Cubaenerg\'{\i}a, Cuba;
Ministry of Education, Youth and Sports of the Czech Republic, Czech Republic;
The Danish Council for Independent Research | Natural Sciences, the VILLUM FONDEN and Danish National Research Foundation (DNRF), Denmark;
Helsinki Institute of Physics (HIP), Finland;
Commissariat \`{a} l'Energie Atomique (CEA) and Institut National de Physique Nucl\'{e}aire et de Physique des Particules (IN2P3) and Centre National de la Recherche Scientifique (CNRS), France;
Bundesministerium f\"{u}r Forschung, Technologie und Raumfahrt (BMFTR) and GSI Helmholtzzentrum f\"{u}r Schwerionenforschung GmbH, Germany;
General Secretariat for Research and Technology, Ministry of Education, Research and Religions, Greece;
National Research, Development and Innovation Office, Hungary;
Department of Atomic Energy Government of India (DAE), Department of Science and Technology, Government of India (DST), University Grants Commission, Government of India (UGC) and Council of Scientific and Industrial Research (CSIR), India;
National Research and Innovation Agency - BRIN, Indonesia;
Istituto Nazionale di Fisica Nucleare (INFN), Italy;
Japanese Ministry of Education, Culture, Sports, Science and Technology (MEXT) and Japan Society for the Promotion of Science (JSPS) KAKENHI, Japan;
Consejo Nacional de Ciencia (CONACYT) y Tecnolog\'{i}a, through Fondo de Cooperaci\'{o}n Internacional en Ciencia y Tecnolog\'{i}a (FONCICYT) and Direcci\'{o}n General de Asuntos del Personal Academico (DGAPA), Mexico;
Nederlandse Organisatie voor Wetenschappelijk Onderzoek (NWO), Netherlands;
The Research Council of Norway, Norway;
Pontificia Universidad Cat\'{o}lica del Per\'{u}, Peru;
Ministry of Science and Higher Education, National Science Centre and WUT ID-UB, Poland;
Korea Institute of Science and Technology Information and National Research Foundation of Korea (NRF), Republic of Korea;
Ministry of Education and Scientific Research, Institute of Atomic Physics, Ministry of Research and Innovation and Institute of Atomic Physics and Universitatea Nationala de Stiinta si Tehnologie Politehnica Bucuresti, Romania;
Ministerstvo skolstva, vyskumu, vyvoja a mladeze SR, Slovakia;
National Research Foundation of South Africa, South Africa;
Swedish Research Council (VR) and Knut \& Alice Wallenberg Foundation (KAW), Sweden;
European Organization for Nuclear Research, Switzerland;
Suranaree University of Technology (SUT), National Science and Technology Development Agency (NSTDA) and National Science, Research and Innovation Fund (NSRF via PMU-B B05F650021), Thailand;
Turkish Energy, Nuclear and Mineral Research Agency (TENMAK), Turkey;
National Academy of  Sciences of Ukraine, Ukraine;
Science and Technology Facilities Council (STFC), United Kingdom;
National Science Foundation of the United States of America (NSF) and United States Department of Energy, Office of Nuclear Physics (DOE NP), United States of America.
In addition, individual groups or members have received support from:
Czech Science Foundation (grant no. 23-07499S), Czech Republic;
FORTE project, reg.\ no.\ CZ.02.01.01/00/22\_008/0004632, Czech Republic, co-funded by the European Union, Czech Republic;
European Research Council (grant no. 950692), European Union;
Deutsche Forschungs Gemeinschaft (DFG, German Research Foundation) ``Neutrinos and Dark Matter in Astro- and Particle Physics'' (grant no. SFB 1258), Germany;
ICSC - National Research Center for High Performance Computing, Big Data and Quantum Computing and FAIR - Future Artificial Intelligence Research, funded by the NextGenerationEU program (Italy).
\end{acknowledgement}

\bibliographystyle{utphys}   
\bibliography{biblio}

\newpage
\appendix
\section{The ALICE Collaboration}
\label{app:collab}
\begin{flushleft} 
\small

I.J.~Abualrob\,\orcidlink{0009-0005-3519-5631}\,$^{\rm 114}$, 
S.~Acharya\,\orcidlink{0000-0002-9213-5329}\,$^{\rm 50}$, 
G.~Aglieri Rinella\,\orcidlink{0000-0002-9611-3696}\,$^{\rm 32}$, 
L.~Aglietta\,\orcidlink{0009-0003-0763-6802}\,$^{\rm 24}$, 
M.~Agnello\,\orcidlink{0000-0002-0760-5075}\,$^{\rm 29}$, 
N.~Agrawal\,\orcidlink{0000-0003-0348-9836}\,$^{\rm 25}$, 
Z.~Ahammed\,\orcidlink{0000-0001-5241-7412}\,$^{\rm 133}$, 
S.~Ahmad\,\orcidlink{0000-0003-0497-5705}\,$^{\rm 15}$, 
I.~Ahuja\,\orcidlink{0000-0002-4417-1392}\,$^{\rm 36}$, 
ZUL.~Akbar$^{\rm 81}$, 
A.~Akindinov\,\orcidlink{0000-0002-7388-3022}\,$^{\rm 139}$
% ...

\section*{Affiliation Notes}

$^{\rm I}$ Deceased\\
$^{\rm II}$ Also at: Max-Planck-Institut fur Physik, Munich, Germany\\
$^{\rm III}$ Also at: Czech Technical University in Prague (CZ)\\
$^{\rm IV}$ Also at: Italian National Agency for New Technologies, Energy and Sustainable Economic Development (ENEA), Bologna, Italy\\
$^{\rm V}$ Also at: Instituto de Fisica da Universidade de Sao Paulo\\
$^{\rm VI}$ Also at: Dipartimento DET del Politecnico di Torino, Turin, Italy\\
$^{\rm VII}$ Also at: Department of Applied Physics, Aligarh Muslim University, Aligarh, India\\
$^{\rm VIII}$ Also at: Institute of Theoretical Physics, University of Wroclaw, Poland\\
$^{\rm IX}$ Also at: Facultad de Ciencias, Universidad Nacional Aut\'{o}noma de M\'{e}xico, Mexico City, Mexico\\

\section*{Collaboration Institutes}

$^{1}$ A.I. Alikhanyan National Science Laboratory (Yerevan Physics Institute) Foundation, Yerevan, Armenia\\
$^{2}$ AGH University of Krakow, Cracow, Poland\\
$^{3}$ Bogolyubov Institute for Theoretical Physics, National Academy of Sciences of Ukraine, Kyiv, Ukraine\\
$^{4}$ Bose Institute, Department of Physics  and Centre for Astroparticle Physics and Space Science (CAPSS), Kolkata, India\\
$^{5}$ California Polytechnic State University, San Luis Obispo, California, United States\\
$^{6}$ Central China Normal University, Wuhan, China\\
$^{7}$ Centro de Aplicaciones Tecnol\'{o}gicas y Desarrollo Nuclear (CEADEN), Havana, Cuba\\
$^{8}$ Centro de Investigaci\'{o}n y de Estudios Avanzados (CINVESTAV), Mexico City and M\'{e}rida, Mexico\\
$^{9}$ Chicago State University, Chicago, Illinois, United States\\
$^{10}$ China Nuclear Data Center, China Institute of Atomic Energy, Beijing, China\\
$^{11}$ China University of Geosciences, Wuhan, China\\
$^{12}$ Chungbuk National University, Cheongju, Republic of Korea\\
$^{13}$ Comenius University Bratislava, Faculty of Mathematics, Physics and Informatics, Bratislava, Slovak Republic\\
$^{14}$ Creighton University, Omaha, Nebraska, United States\\
$^{15}$ Department of Physics, Aligarh Muslim University, Aligarh, India\\
$^{16}$ Department of Physics, Pusan National University, Pusan, Republic of Korea\\
$^{17}$ Department of Physics, Sejong University, Seoul, Republic of Korea\\
$^{18}$ Department of Physics, University of California, Berkeley, California, United States\\
$^{19}$ Department of Physics, University of Oslo, Oslo, Norway\\
$^{20}$ Department of Physics and Technology, University of Bergen, Bergen, Norway\\
$^{21}$ Dipartimento di Fisica, Universit\`{a} di Pavia, Pavia, Italy\\
$^{22}$ Dipartimento di Fisica dell'Universit\`{a} and Sezione INFN, Cagliari, Italy\\
$^{23}$ Dipartimento di Fisica dell'Universit\`{a} and Sezione INFN, Trieste, Italy\\
$^{24}$ Dipartimento di Fisica dell'Universit\`{a} and Sezione INFN, Turin, Italy\\
$^{25}$ Dipartimento di Fisica e Astronomia dell'Universit\`{a} and Sezione INFN, Bologna, Italy\\
$^{26}$ Dipartimento di Fisica e Astronomia dell'Universit\`{a} and Sezione INFN, Catania, Italy\\
$^{27}$ Dipartimento di Fisica e Astronomia dell'Universit\`{a} and Sezione INFN, Padova, Italy\\
$^{28}$ Dipartimento di Fisica `E.R.~Caianiello' dell'Universit\`{a} and Gruppo Collegato INFN, Salerno, Italy\\
$^{29}$ Dipartimento DISAT del Politecnico and Sezione INFN, Turin, Italy\\
$^{30}$ Dipartimento di Scienze MIFT, Universit\`{a} di Messina, Messina, Italy\\
$^{31}$ Dipartimento Interateneo di Fisica `M.~Merlin' and Sezione INFN, Bari, Italy\\
$^{32}$ European Organization for Nuclear Research (CERN), Geneva, Switzerland\\
$^{33}$ Faculty of Electrical Engineering, Mechanical Engineering and Naval Architecture, University of Split, Split, Croatia\\
$^{34}$ Faculty of Nuclear Sciences and Physical Engineering, Czech Technical University in Prague, Prague, Czech Republic\\
$^{35}$ Faculty of Physics, Sofia University, Sofia, Bulgaria\\
$^{36}$ Faculty of Science, P.J.~\v{S}af\'{a}rik University, Ko\v{s}ice, Slovak Republic\\
$^{37}$ Faculty of Technology, Environmental and Social Sciences, Bergen, Norway\\
$^{38}$ Frankfurt Institute for Advanced Studies, Johann Wolfgang Goethe-Universit\"{a}t Frankfurt, Frankfurt, Germany\\
$^{39}$ Fudan University, Shanghai, China\\
$^{40}$ Gangneung-Wonju National University, Gangneung, Republic of Korea\\
$^{41}$ Gauhati University, Department of Physics, Guwahati, India\\
$^{42}$ Helmholtz-Institut f\"{u}r Strahlen- und Kernphysik, Rheinische Friedrich-Wilhelms-Universit\"{a}t Bonn, Bonn, Germany\\
$^{43}$ Helsinki Institute of Physics (HIP), Helsinki, Finland\\
$^{44}$ High Energy Physics Group,  Universidad Aut\'{o}noma de Puebla, Puebla, Mexico\\
$^{45}$ Horia Hulubei National Institute of Physics and Nuclear Engineering, Bucharest, Romania\\
$^{46}$ HUN-REN Wigner Research Centre for Physics, Budapest, Hungary\\
$^{47}$ Indian Institute of Technology Bombay (IIT), Mumbai, India\\
$^{48}$ Indian Institute of Technology Indore, Indore, India\\
$^{49}$ INFN, Laboratori Nazionali di Frascati, Frascati, Italy\\
$^{50}$ INFN, Sezione di Bari, Bari, Italy\\
$^{51}$ INFN, Sezione di Bologna, Bologna, Italy\\
$^{52}$ INFN, Sezione di Cagliari, Cagliari, Italy\\
$^{53}$ INFN, Sezione di Catania, Catania, Italy\\
$^{54}$ INFN, Sezione di Padova, Padova, Italy\\
$^{55}$ INFN, Sezione di Pavia, Pavia, Italy\\
$^{56}$ INFN, Sezione di Torino, Turin, Italy\\
$^{57}$ INFN, Sezione di Trieste, Trieste, Italy\\
$^{58}$ Inha University, Incheon, Republic of Korea\\
$^{59}$ Institute for Gravitational and Subatomic Physics (GRASP), Utrecht University/Nikhef, Utrecht, Netherlands\\
$^{60}$ Institute of Experimental Physics, Slovak Academy of Sciences, Ko\v{s}ice, Slovak Republic\\
$^{61}$ Institute of Physics, Homi Bhabha National Institute, Bhubaneswar, India\\
$^{62}$ Institute of Physics of the Czech Academy of Sciences, Prague, Czech Republic\\
$^{63}$ Institute of Space Science (ISS), Bucharest, Romania\\
$^{64}$ Institut f\"{u}r Kernphysik, Johann Wolfgang Goethe-Universit\"{a}t Frankfurt, Frankfurt, Germany\\
$^{65}$ Instituto de Ciencias Nucleares, Universidad Nacional Aut\'{o}noma de M\'{e}xico, Mexico City, Mexico\\
$^{66}$ Instituto de F\'{i}sica, Universidade Federal do Rio Grande do Sul (UFRGS), Porto Alegre, Brazil\\
$^{67}$ Instituto de F\'{\i}sica, Universidad Nacional Aut\'{o}noma de M\'{e}xico, Mexico City, Mexico\\
$^{68}$ iThemba LABS, National Research Foundation, Somerset West, South Africa\\
$^{69}$ Jeonbuk National University, Jeonju, Republic of Korea\\
$^{70}$ Johann-Wolfgang-Goethe Universit\"{a}t Frankfurt Institut f\"{u}r Informatik, Fachbereich Informatik und Mathematik, Frankfurt, Germany\\
$^{71}$ Korea Institute of Science and Technology Information, Daejeon, Republic of Korea\\
$^{72}$ Laboratoire de Physique Subatomique et de Cosmologie, Universit\'{e} Grenoble-Alpes, CNRS-IN2P3, Grenoble, France\\
$^{73}$ Lawrence Berkeley National Laboratory, Berkeley, California, United States\\
$^{74}$ Lund University Department of Physics, Division of Particle Physics, Lund, Sweden\\
$^{75}$ Nagasaki Institute of Applied Science, Nagasaki, Japan\\
$^{76}$ Nara Women{'}s University (NWU), Nara, Japan\\
$^{77}$ National and Kapodistrian University of Athens, School of Science, Department of Physics , Athens, Greece\\
$^{78}$ National Centre for Nuclear Research, Warsaw, Poland\\
$^{79}$ National Institute of Science Education and Research, Homi Bhabha National Institute, Jatni, India\\
$^{80}$ National Nuclear Research Center, Baku, Azerbaijan\\
$^{81}$ National Research and Innovation Agency - BRIN, Jakarta, Indonesia\\
$^{82}$ Niels Bohr Institute, University of Copenhagen, Copenhagen, Denmark\\
$^{83}$ Nikhef, National institute for subatomic physics, Amsterdam, Netherlands\\
$^{84}$ Nuclear Physics Group, STFC Daresbury Laboratory, Daresbury, United Kingdom\\
$^{85}$ Nuclear Physics Institute of the Czech Academy of Sciences, Husinec-\v{R}e\v{z}, Czech Republic\\
$^{86}$ Oak Ridge National Laboratory, Oak Ridge, Tennessee, United States\\
$^{87}$ Ohio State University, Columbus, Ohio, United States\\
$^{88}$ Physics department, Faculty of science, University of Zagreb, Zagreb, Croatia\\
$^{89}$ Physics Department, Panjab University, Chandigarh, India\\
$^{90}$ Physics Department, University of Jammu, Jammu, India\\
$^{91}$ Physics Program and International Institute for Sustainability with Knotted Chiral Meta Matter (WPI-SKCM$^{2}$), Hiroshima University, Hiroshima, Japan\\
$^{92}$ Physikalisches Institut, Eberhard-Karls-Universit\"{a}t T\"{u}bingen, T\"{u}bingen, Germany\\
$^{93}$ Physikalisches Institut, Ruprecht-Karls-Universit\"{a}t Heidelberg, Heidelberg, Germany\\
$^{94}$ Physik Department, Technische Universit\"{a}t M\"{u}nchen, Munich, Germany\\
$^{95}$ Politecnico di Bari and Sezione INFN, Bari, Italy\\
$^{96}$ Research Division and ExtreMe Matter Institute EMMI, GSI Helmholtzzentrum f\"ur Schwerionenforschung GmbH, Darmstadt, Germany\\
$^{97}$ Saga University, Saga, Japan\\
$^{98}$ Saha Institute of Nuclear Physics, Homi Bhabha National Institute, Kolkata, India\\
$^{99}$ School of Physics and Astronomy, University of Birmingham, Birmingham, United Kingdom\\
$^{100}$ Secci\'{o}n F\'{\i}sica, Departamento de Ciencias, Pontificia Universidad Cat\'{o}lica del Per\'{u}, Lima, Peru\\
$^{101}$ Stefan Meyer Institut f\"{u}r Subatomare Physik (SMI), Vienna, Austria\\
$^{102}$ SUBATECH, IMT Atlantique, Nantes Universit\'{e}, CNRS-IN2P3, Nantes, France\\
$^{103}$ Sungkyunkwan University, Suwon City, Republic of Korea\\
$^{104}$ Suranaree University of Technology, Nakhon Ratchasima, Thailand\\
$^{105}$ Technical University of Ko\v{s}ice, Ko\v{s}ice, Slovak Republic\\
$^{106}$ The Henryk Niewodniczanski Institute of Nuclear Physics, Polish Academy of Sciences, Cracow, Poland\\
$^{107}$ The University of Texas at Austin, Austin, Texas, United States\\
$^{108}$ Universidad Aut\'{o}noma de Sinaloa, Culiac\'{a}n, Mexico\\
$^{109}$ Universidade de S\~{a}o Paulo (USP), S\~{a}o Paulo, Brazil\\
$^{110}$ Universidade Estadual de Campinas (UNICAMP), Campinas, Brazil\\
$^{111}$ Universidade Federal do ABC, Santo Andre, Brazil\\
$^{112}$ Universitatea Nationala de Stiinta si Tehnologie Politehnica Bucuresti, Bucharest, Romania\\
$^{113}$ University of Derby, Derby, United Kingdom\\
$^{114}$ University of Houston, Houston, Texas, United States\\
$^{115}$ University of Jyv\"{a}skyl\"{a}, Jyv\"{a}skyl\"{a}, Finland\\
$^{116}$ University of Kansas, Lawrence, Kansas, United States\\
$^{117}$ University of Liverpool, Liverpool, United Kingdom\\
$^{118}$ University of Science and Technology of China, Hefei, China\\
$^{119}$ University of South-Eastern Norway, Kongsberg, Norway\\
$^{120}$ University of Tennessee, Knoxville, Tennessee, United States\\
$^{121}$ University of the Witwatersrand, Johannesburg, South Africa\\
$^{122}$ University of Tokyo, Tokyo, Japan\\
$^{123}$ University of Tsukuba, Tsukuba, Japan\\
$^{124}$ Universit\"{a}t M\"{u}nster, Institut f\"{u}r Kernphysik, M\"{u}nster, Germany\\
$^{125}$ Universit\'{e} Clermont Auvergne, CNRS/IN2P3, LPC, Clermont-Ferrand, France\\
$^{126}$ Universit\'{e} de Lyon, CNRS/IN2P3, Institut de Physique des 2 Infinis de Lyon, Lyon, France\\
$^{127}$ Universit\'{e} de Strasbourg, CNRS, IPHC UMR 7178, F-67000 Strasbourg, France, Strasbourg, France\\
$^{128}$ Universit\'{e} Paris-Saclay, Centre d'Etudes de Saclay (CEA), IRFU, D\'{e}partment de Physique Nucl\'{e}aire (DPhN), Saclay, France\\
$^{129}$ Universit\'{e}  Paris-Saclay, CNRS/IN2P3, IJCLab, Orsay, France\\
$^{130}$ Universit\`{a} degli Studi di Foggia, Foggia, Italy\\
$^{131}$ Universit\`{a} del Piemonte Orientale, Vercelli, Italy\\
$^{132}$ Universit\`{a} di Brescia, Brescia, Italy\\
$^{133}$ Variable Energy Cyclotron Centre, Homi Bhabha National Institute, Kolkata, India\\
$^{134}$ Warsaw University of Technology, Warsaw, Poland\\
$^{135}$ Wayne State University, Detroit, Michigan, United States\\
$^{136}$ Yale University, New Haven, Connecticut, United States\\
$^{137}$ Yildiz Technical University, Istanbul, Turkey\\
$^{138}$ Yonsei University, Seoul, Republic of Korea\\
$^{139}$ Affiliated with an institute formerly covered by a cooperation agreement with CERN\\
$^{140}$ Affiliated with an international laboratory covered by a cooperation agreement with CERN.\\

\end{flushleft} 
\end{document}